\documentclass[11pt]{article}
\usepackage{fancyhdr}
\usepackage{isomath}
\usepackage{amsmath}
\usepackage{amsbsy}
\usepackage{amssymb}
\usepackage{amscd}
\usepackage{amsfonts}
\usepackage{array}
\usepackage{graphicx}
\usepackage{verbatim}
\usepackage{euscript}
\usepackage{alltt}
\usepackage{stmaryrd}
\usepackage{relsize}
\usepackage{enumitem}
\usepackage[lofdepth,lotdepth]{subfig}
\usepackage[numbers]{natbib}
\usepackage{caption}
\usepackage{float}
\usepackage{multirow}
\usepackage{epsfig}
\usepackage{color}
\usepackage{soul}
\usepackage{ulem}
\usepackage{comment}

\DeclareGraphicsExtensions{.eps,.pdf,.png}
\usepackage{amsmath}
\usepackage{amsbsy}
\usepackage{amssymb}
\usepackage{amscd}
\usepackage{amsfonts}

\newcommand{\bs}{\boldsymbol}

\newcommand{\bfa}{{\mathbold a}}
\newcommand{\bfb}{{\mathbold b}}
\newcommand{\bfc}{{\mathbold c}}
\newcommand{\bfd}{{\mathbold d}}
\newcommand{\bfe}{{\mathbold e}}
\newcommand{\bff}{{\mathbold f}}

\newcommand{\bfn}{{\mathbold n}}

\newcommand{\bfv}{{\mathbold v}}

\newcommand{\bfx}{{\mathbold x}}

\newcommand{\bfA}{{\mathbold A}}
\newcommand{\bfB}{{\mathbold B}}

\newcommand{\bfE}{{\mathbold E}}
\newcommand{\bfF}{{\mathbold F}}

\newcommand{\bfL}{{\mathbold L}}

\newcommand{\bfT}{{\mathbold T}}

\newcommand{\bfV}{{\mathbold V}}
\newcommand{\bfW}{{\mathbold W}}
\newcommand{\bfX}{{\mathbold X}}

\newcommand{\beq}{\begin{equation}}
\newcommand{\eeq}{\end{equation}}
\newcommand{\beqs}{\begin{eqnarray}}
\newcommand{\eeqs}{\end{eqnarray}}
\newcommand{\beql}{\begin{equation} \label}

\newcommand{\bfchi}{\mathbold{\chi}}

\newcommand{\bfalpha}{\mathbold{\alpha}}

\newcommand{\bfzero}{\mathbf{0}}

\newcommand{\grad}{\mathop{\rm grad}\nolimits}
\newcommand{\divergence}{\mathop{\rm div}\nolimits}
\newcommand{\curl}{\mathop{\rm curl}\nolimits}

\usepackage[margin=1in]{geometry}

\date{}
\begin{document}
\title{Mechanics of micropillar confined thin film plasticity}
\author{Abhishek Arora\thanks{Department of Civil \& Environmental Engineering, Carnegie Mellon University, Pittsburgh, PA 15213} $\qquad$  Rajat Arora\thanks{Siemens Corporate Technology, Princeton, NJ} $\qquad$  Amit Acharya\thanks{Department of Civil \& Environmental Engineering, and Center for Nonlinear Analysis, Carnegie Mellon University, Pittsburgh, PA 15213, email: acharyaamit@cmu.edu.}}
\maketitle

\begin{abstract}
 \noindent Micropillar compression experiments probing size effects in confined plasticity of metal thin films, including the indirect imposition of `canonical' simple shearing boundary conditions, show dramatically different responses in compression and shear of the film. The Mesoscale Field Dislocation Mechanics (MFDM) model is confronted with this set of experimental observations and shown to be capable of modeling such behavior, without any ad-hoc modification to the basic structure of the theory (including boundary conditions), or the use of extra fitting parameters. This is a required theoretical advance in the current state-of-the art of strain gradient plasticity models. It is also shown that significantly different inhomogeneous fields can display qualitatively similar size effect trends in overall agreement with the experimental results. The (plastic) Swift and (elastic) Poynting finite deformation effects are also demonstrated.
\end{abstract}

\section{Introduction}
Size-effects in confined plasticity of metal thin films sandwiched in ceramic micropillars have been demonstrated in the work of Meng and co-workers \cite{mu2014thickness, mu2016dependence, mu2014micro}.  Shear failure testing of the interfacial regions of CrN/Cu/Si and CrN/Ti/Si ceramic-coating/metal-adhesion layer/substrate systems through instrumented compression of cylindrical micropillars was reported in  \cite{zhang2017mechanical}. The last thirty years have seen intense world-wide activity in modeling size effects in metal plasticity at the microscale, initiated by the Strain Gradient Plasticity (SGP) work of Fleck, Hutchinson and co-workers \cite{fleck1994strain}; the study of length-scale effects in plasticity was initiated earlier by Aifantis \cite{aifantis1987physics} and co-workers. The confined plasticity results in \cite{mu2016dependence}, however, have not been successfully modeled by SGP as pointed out in \cite{mu2016dependence}, without rather drastic modifications to the structure of the theory and introducing extra fitting parameters \cite{kuroda2019nonuniform,kuroda2019simple,kuroda2021constraint}. The aim of our work is to report on the reasonably successful modeling of the experiments in \cite{mu2016dependence} with the Mesoscale Field Dislocation Mechanics theory \cite{acharya2006size,arora2020dislocation,arora2020finite,arora2020unification}, and to provide a mechanistic understanding of the observed effects within the idealization of the model. We also analyze the mechanics of local fields and other interesting results bearing on historically important observed effects in the large deformation of elastic-plasic materials.
 
Conventional plasticity models do not have a material length scale and produce size-independent response (for homogeneous materials). In SGP theory, the material response is assumed to depend on both the plastic strain and its spatial gradient \cite{ gurtin2000plasticity,fleck2001reformulation,gudmundson2004unified}, and the work conjugate of the plastic strain gradient is interpreted to be a `microscopic stress,' of unspecified physical origin. SGP theories predict a much stronger dependence on the  film thickness under nominal simple shearing conditions compared to experimental observations \cite{mu2016dependence,mu2014thickness}. This prompted Kuroda and Needleman \cite{kuroda2019simple} to introduce an ad-hoc modification to the constrained boundary condition specification of SGP theory where it is assumed that a threshold exists on the magnitude of the plastic strain gradient at boundaries. Above this threshold, the constrained boundary condition is released and plastic straining is allowed at the boundaries - as is clear, this threshold is not a material parameter, and it is not clear what its validity is, and how it is to be determined, in general modes of loading. Kuroda et al.~\cite{kuroda2021constraint} use this boundary condition in a finite deformation setting along with extensive fitting of their model parameters to the data of \cite{mu2016dependence} (including the classical work-hardening modulus and the initial yield stress) to produce results in accord with the experimental observations on compression of ceramic-metal thin film sandwich micropillars. Another effort to address this shortcoming of SGP theory is that of Dahlberg and Ortiz \cite{dahlberg2019fractional} who introduce a fractional derivative of the effective plastic strain in the material response (a fractional SGP theory), whose fractional exponent is intended as a fitting parameter to recover the experimentally observed scaling of shear stress with layer thickness. Arora and Acharya \cite{arora2020unification} used the finite deformation implementation \citep{arora2020finite} of MFDM theory, without any special fitting beyond the use of generic material parameters used in MFDM simulations of polycrystalline metals, to model constrained shearing of a thin film. They recovered the observed size-effect trend in \cite{mu2016dependence} corresponding to the $45^\circ$-oriented thin film in the micropillar experiments.

\begin{figure}[htbp]
    \centering
    \subfloat[][$90^\circ$ orientation]{
    \includegraphics[scale=0.3]{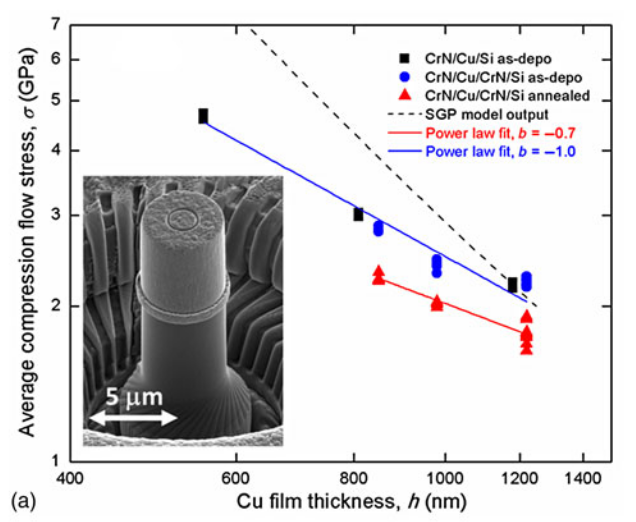}}
    \subfloat[][$45^\circ$ orientation]{
    \includegraphics[scale=0.3]{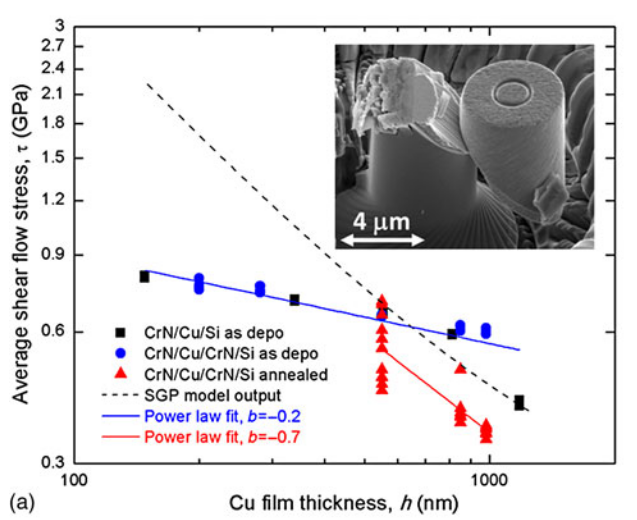}}
    \caption{ Micropillar experimental specimen and size effect results in both $90^{\circ}$ and $45^{\circ}$ orientations (Figures reprinted from \cite{mu2016dependence} with permission from \textit{Springer Nature}).}
    \label{fig:experiment}
\end{figure}

\begin{figure}[htbp]
    \centering
    \subfloat[][$90^\circ$ orientation]{
    \includegraphics[scale=0.4]{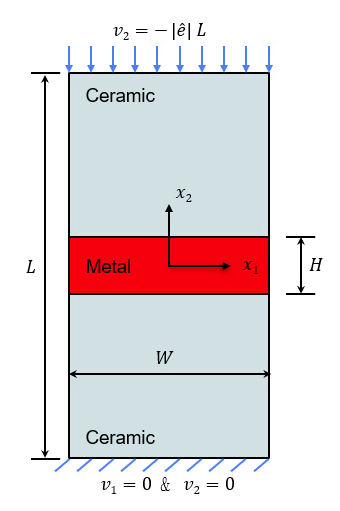}}
    \subfloat[][$45^\circ$ orientation]{
    \includegraphics[scale=0.4]{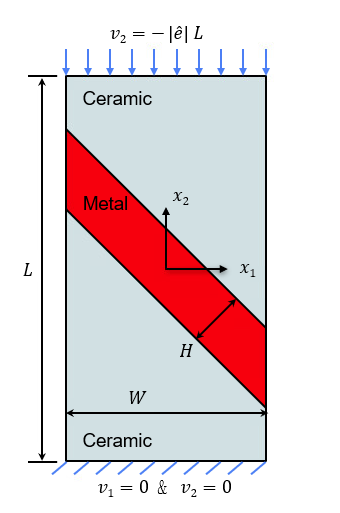}}
    \caption{Schematic of idealized plane strain model of micropillar compression experiments.}
    \label{fig:schematic}
\end{figure}

Moving beyond the simple thin film-only geometry, in this work we use the MFDM framework to study the compression experiments of \cite{mu2016dependence}, involving metal thin films sandwiched in ceramic micropillars, within a plane-strain idealization. As in the experiments, both $45^\circ$ and $90^\circ$ oriented thin films are considered. The experimental specimen and the corresponding size effect results are shown in Fig.~\ref{fig:experiment}, and we compare the size effect results from our simulations to results for the as-deposited samples. The metal layer is sandwiched between two ceramic blocks above and below it, and the whole composite is put under compression. The schematic of pillars with the thin film in the $90^{\circ}$ and $45^{\circ}$ orientations is shown in Fig.\ \ref{fig:schematic}. As explained in detail in \cite{mu2016dependence}, compression of the micropillar with the film in the $45^{\circ}$ orientation results in a nominal simple shearing boundary condition imposed on it; compression with the film in the $90^{\circ}$ orientation results in compression boundary conditions on the film, with restrained lateral movement at its top and bottom boundaries. The two configurations result in dramatically different observations of size effects, which we recover.
 
This paper is organised as follows: the following paragraph contains some notational details. Section \ref{theory} recalls the MFDM governing equations, boundary, and initial conditions. For details on the physical basis of the model and the computational framework, the interested reader is referred to  \cite{arora2020dislocation,arora2020finite,arora2019computational}. Section \ref{results_and_discussion} shows the scaling obtained for applied nominal stress with thin film thickness, for both $90^{\circ}$ and $45^{\circ}$ orientations of the film, and provides a mechanistic explanation for the observations. It is also shown that a free-standing film under simple shear produces normal stress, as observed in metals bars under torsion \cite{billington1977non}, and the Poynting effect in non-linear elastic solids \cite{poynting1909pressure}. Section \ref{conclusion} contains some concluding remarks.

Vectors and tensors are represented by boldface lower and upper-case letters. The action of a second-order tensor $\bfA$ on a vector $\bfb$ is denoted by $\bfA \bfb$. The inner product of two vectors is denoted by $\bfa \cdot \bfb$, while the inner product of two second-order tensors is denoted by $\bfA:\bfB$. A rectangular Cartesian coordinate system is invoked for ambient space and all (vector) tensor components are expressed with respect to the basis of this coordinate system. $(\cdot)_{,i}$ denotes the partial derivative of the quantity $(\cdot)$ w.r.t.\ the $x_i$ coordinate of this coordinate system. $\bfe_i$ denotes the unit vector in the $x_i$ direction.  The time derivative of a quantity is denoted by $\dot{(\cdot)}$. Einstein's summation convection is always implied unless mentioned otherwise. The symbols $\grad$, $\divergence$, and $\curl$ denote the gradient, divergence, and curl on the current configuration. For a second order tensor $\bfA$, vectors $\bfv$, $\bfa$, and $\bfc$, a spatially constant vector field $\bfb$, the operations of $\divergence$, $\curl$, and cross-product of a tensor $(\times)$ with a vector are defined as follows: 
\begin{subequations}
\begin{align*}
    \left( \divergence \bfA \right) \cdot \bfb = \divergence \left( \bfA^T \bfb \right),& \qquad \forall \:\: \bfb \\
    \bfb \cdot \left( \curl \bfA \right) \bfc = \left[ \curl\left(\bfA^T \bfb \right)\right] \cdot \bfc,& \qquad \forall \: \: \bfb, \bfc \\
    \bfc \cdot \left( \bfA \times \bfv \right) \bfa = \left[ \left(\bfA^T \bfc \right) \times \bfv \right] \cdot \bfa,& \qquad \forall \: \: \bfa, \bfc.
\end{align*}
\end{subequations}
In rectangular Cartesian coordinates, these are denoted by
\begin{subequations}
\begin{align*}
    \left( \divergence \bfA \right)_i =  A_{ij,j}, \quad
    \left( \curl \bfA \right)_{ri} = \varepsilon_{ijk} A_{rk,j}, \quad
    \left( \bfA \times \bfv \right)_{ri} = \varepsilon_{ijk} A_{rj} v_{k},
\end{align*}
\end{subequations}
where $\varepsilon_{ijk}$ are the components of the third order alternating tensor $\bfX$. 


\section{Theory}\label{theory}
The governing equations for MFDM \citep{arora2020finite} are as follows:
\begin{subequations}
\begin{gather}
     \mathring{\bs{\alpha}} \equiv (\divergence(\bs{v})) \bs{\alpha} + \dot{\bs{\alpha}} - \bs{\alpha} \bfL^T = - \curl (\bs{\alpha} \times \bfV +  \bfL^p) \label{alpha_conservation} \\
    \begin{array}{c} 
      \bfW = \bs{\chi} + \grad \bff, \: \bs{F}^e = \bs{W}^{-1} \\
      \curl \bs{\chi} = - \bs{\alpha}  \\
    \divergence \bs{\chi}  = \bs{0}          
    \end{array} \Bigg \} \label{incompatibility} \\ 
    \divergence \left( \grad \dot{\bs{f}} \right) = \divergence \left( \bs{\alpha} \times \bs{V} +  \bfL^p - \dot{\bs{\chi}} - \bs{\chi} \bs{L} \right) \label{plastic_eq} \\
    \divergence \left[ \bfT (\bfW) \right] = \bigg\{ \begin{array}{l}
    \bs{0} \quad \text{quasistatic} \\
    \rho \dot{\bfv} \quad \text{dynamic}.
    \end{array}  \label{equilibrium}
\end{gather}
\end{subequations}
Here, $\bs{F}^e$ is the elastic distortion tensor and $\bs{W}$ is the its inverse, $\bs{\chi}$ is the incompatible part of $\bfW$, $\bff$ is the plastic position vector, and $\grad \bff$ is the compatible part of $\bfW$. $\bfalpha$ is the dislocation density tensor, $\bfv$ is the material velocity, $\bfL = \grad \bfv$ is the velocity gradient, $\bfT$ is the (symmetric) Cauchy stress tensor, while $\bfL^p$ is the additional meso-scale field which represents the averaged rate of plastic straining due to all dislocations that cannot be represented by $\bfalpha \times \bfV$, where both fields in the product represent space-time running averages. The fields in the MFDM framework are the running space-time averages of corresponding fields of FDM theory.

To close the above set of equations, constitutive statements for $\bs{V}, \bs{L}^p, \bs{T}$ are used consistent with the mechanical dissipation being non-negative \citep{arora2020dislocation}.

In the present work, plastic straining in accordance with $\mathrm{\textit{J}_2}$ plasticity theory is augmented with the regularizing effects of an idealization of microscopic dislocation core energy to define $\bfL^p$. 

Initial conditions for $\bfalpha, \bff$ and boundary conditions for $\bfalpha, \bff, \bfchi$ and $\bfv$ are specfied for well-set evolution.

\subsection{Boundary conditions}
\begin{itemize}
    \item The $\bs{\alpha}$ evolution equation has a convective boundary condition of the form $(\bfalpha \times \bfV + \bfL^p) \times \bfn = \bs{\Phi} $, where $\bs{\Phi}$ is a second order tensor valued function of time and position on the boundary characterising the flux of dislocations at the surface satisfying the constraint $\bs{\Phi} \bfn = \bfzero$. Here, $\bfn$ is the outward unit normal field on the boundary.
    
    There are two ways in which the boundary condition is specified: (a) \textit{Constrained}:  $\bs{\Phi} (\bfx , t) = \bfzero$ at a point $\bfx$ on the boundary for all times, which ensures that there is no outflow of dislocations at that point of the boundary, and only parallel motion along the boundary is allowed. (b) \textit{Unconstrained}: A less restrictive boundary condition where $\hat{\bfL}^p \times \bfn$ is simply evaluated at the boundary (akin to an outflow condition), along with the specification of dislocation flux $\bfalpha (\bfV \cdot \bfn)$ on the inflow part of the boundary. Additionally, for all calculations presented in this paper $(\curl \bfalpha \times \bfn) = \bf0$ is imposed, a particular specification of a boundary condition that arises from simple mathematical modeling of the manifestation of dislocation core energy at the mesoscale.
    \item For the incompatibility equation, $\bs{\chi} \bs{n} = \bfzero$ is applied on the outer boundary of the domain, which along with the system \eqref{incompatibility} ensures that $\bfchi$ vanishes when $\bfalpha$ is zero in the entire domain. 
    \item The $\bs{f}$ evolution equation requires a Neumann boundary condition i.e., $(\grad \dot{\bff}) \bfn = (\bfalpha \times \bfV + \bfL^p - \dot{\bfchi} - \bfchi \bfL ) \bfn$ on the outer boundary of the domain.
    \item The material velocity boundary conditions are applied based on the loading type, which is discussed later in Section \ref{results_and_discussion}. 
\end{itemize}

\subsection{Initial conditions}
\begin{itemize}
    \item The initial condition $\bs{\alpha} (\bs{x}, 0) = \bs{0}$ is assumed for all sample sizes. 
    \item In general, the initial condition for $\bs{f}$ is obtained by solving for $\bfchi$ from the incompatibility equation and solving for $\bs{f}$ from the equilibrium equation, for prescribed $\bs{\alpha}$ on the given initial  configuration. We refer to this scheme as the elastic theory of continuously distributed dislocations (ECDD). For the inital conditions on $\bfalpha$ considered above, this step is trivial, with $\bff = \bfX$, where $\bfX$ is the position field on the initial configuration.
    \item The model admits an arbitrary specification of $\dot{\bff}$ at a point to uniquely evolve $\bff$ using \eqref{plastic_eq} in time, and this rate is prescribed to vanish.
\end{itemize}

\subsection{Constitutive relations}
Constitutive relations in MFDM are required for the stress $\bfT$, the plastic distortion rate $\bfL^p$, and the dislocation velocity $\bfV$. The details of the thermodynamically consistent constitutive formulations can be found in Sec.~3.1 of \cite{arora2020dislocation}. Table \ref{tab:constitutive_relation_T} presents the constitutive relation for Cauchy stress and mesoscopic core energy density for the material. Tables \ref{tab:constitutive_relation_Lp} and \ref{tab:constitutive_relation_V} show the constitutive relations for plastic distortion rate and dislocation velocity, respectively. Table \ref{tab:constitutive_relation_g} shows the evolution equation for material strength. 

{\renewcommand{\arraystretch}{1.8}
\begin{table}[htbp]
    \centering
    \begin{tabular}{l c}
    \hline 
        Saint-Venant-Kirchhoff Material & 
            \(\displaystyle 
                \phi(\bfW) = \frac{1}{2 \rho^*} \bfE^e : \mathbb{C} : \bfE^e,  \quad \bfT = \bfF^e [\mathbb{C} : \bfE^e] {\bfF^e}^T 
              \) \\    \hline 
         Core energy density &
          \(\displaystyle
              \Upsilon (\bfalpha) = \frac{1}{2 \rho^*} \epsilon \bfalpha : \bfalpha
          \) \\ 
    \hline
    \end{tabular}
    \caption{Constitutive relations for Cauchy stress and core energy density.}
    \label{tab:constitutive_relation_T}
\end{table}}

{\renewcommand{\arraystretch}{1.8}
\begin{table}[htbp]
    \centering
    \begin{tabular}{l c}
    \hline 
        \multirow{2}{6em}{$\mathrm{\textit{J}_2}$ plasticity}  & 
            \(\displaystyle
                \hat{\bfL}^p = \hat{\gamma} \bfW \frac{\bfT^{'}}{|\bfT^{'}|}; \quad 
                \hat{\gamma} = \hat{\gamma}_0 \left( \frac{|\bfT^{'}|}{\sqrt{2} g} \right)^{\frac{1}{m}}
            \) \\ 
            &  \(\displaystyle\bfL^p = \hat{\bfL}^p + l^2 \hat{\gamma} \curl \bfalpha \) \label{lp_equation}\\
    \hline
    \end{tabular}
    \caption{Constitutive relations for plastic strain rate $\bfL^p$.}
    \label{tab:constitutive_relation_Lp}
\end{table}}

{\renewcommand{\arraystretch}{1.8}
\begin{table}[htbp]
    \centering
    \begin{tabular}{c c c}
    \hline 
        \(\displaystyle T^{'}_{ij} = T_{ij} - \frac{T_{mm}}{3} \delta_{ij}\);
        &  \(\displaystyle
            a_i = \frac{1}{3} T_{mm} \varepsilon_{ijk} F^{e}_{jp} \alpha_{pk};
            \)    
            &  \(\displaystyle c_l = \varepsilon_{ijk} T^{'}_{jr} F^{e}_{rp} \alpha_{pk} \)  \\
    \( \displaystyle\bfd = \bfc - \left( \bfc - \frac{\bfa}{|\bfa|} \right) \frac{\bfa}{|\bfa|}; \) 
    & \(\displaystyle \bfV = \zeta \frac{\bfd}{|\bfd|} ; \)  
    & \(\displaystyle \zeta = \left(\frac{\mu}{g}\right)^2 \eta^2 b \hat{\gamma}\) \\
    \hline
    \end{tabular}
    \caption{Constitutive relations for dislocation velocity $\bfV$.}
    \label{tab:constitutive_relation_V}
\end{table}}

{\renewcommand{\arraystretch}{1.8}
\begin{table}[H]
    \centering
    \begin{tabular}{c c}
    \hline 
    \( \displaystyle \dot{g} = h(\bs{\alpha}, g) \left(|\bs{F}^e \bs{\alpha} \times \bs{V}| + \hat{\gamma}\right); \) 
    & \( \displaystyle h(\bs{\alpha}, g) =  \frac{\mu^2 \eta^2 b}{2 (g-g_0)} k_{0} |\bs{\alpha}| + \Theta_{0} \left(\frac{g_s-g}{g_s - g_0} \right) \)  \\
    \hline
    \end{tabular}
    \caption{Constitutive relations for material strength $g$.}
    \label{tab:constitutive_relation_g}
\end{table}}
The physical meanings of the material parameters in our model are: $\mu$ is the shear modulus, $\hat{\gamma}_0$ is the reference strain rate, $m$ is the material rate sensitivity, $\eta$ is a non-dimensional material constant in the empirical Taylor relationship for macroscopic strength vs dislocation density, $b$ is the Burgers vector magnitude of a full dislocation in the crystalline material, $g_0$ is the  initial strength (initial yield stress in shear), $g_s$ is the  saturation strength, $\Theta_0$ is the Stage II hardening rate, $k_0$ (non-dimensional) characterizes the hardening rate due to geometrically necessary dislocations (GNDs), and $l$ is a material length related to the gross modeling of mesoscale effects of dislocation core energy, and is defined as $l^2 = \epsilon/g_0$.

All parameters in our model, except $k_0$ and $l$, are part of the constitutive structure of well-accepted models of classical plasticity theory. The parameter $k_0$ was introduced in \cite{acharya2000grain}. The length scale $l$ simply controls the refinement of the GND microstructure and does not play a physically significant role in our results.

\section{Results and discussion} \label{results_and_discussion}
With reference to Fig.~\ref{fig:schematic}, all micropillar compression simulations are performed on initial domain sizes of $5\mu m \times 10 \mu m$, containing polycrystalline Cu thin films in $45^\circ$ and $90^\circ$ orientations. Four (initial) thin film thicknesses of $0.5\mu m$, $0.8\mu m$, $1.0\mu m$, and $1.2 \mu m$ are considered in both orientations. Simulations of thin films with the same thicknesses and of $5 \mu m$ width in free-standing configurations are also performed, under compression and shear loading. The nominal compression loading rate is $|\hat{e}| = 0.001\, s^{-1}$ for both micropillar and free-standing film configurations, and the simple shear loading rate is $\hat{\Gamma} = 0.001 \, s^{-1}$, for the free-standing films. At any time $t$, the nominal compressive strain is $|e| = |\hat{e}| t$, while the nominal shear strain is $\Gamma  = \hat{\Gamma} t$. 

For the micropillar simulations, the interfaces between the thin film and the ceramic blocks are assumed to be plastically unconstrained. For the free-standing films, the top/bottom boundaries are plastically constrained, while the left/right boundaries are plastically unconstrained.

The boundary conditions for material velocity for the micropillar simulations are as follows: 
\begin{itemize}
    \item $v_1 = v_2 = 0$ at the bottom boundary of the domain.
    \item $v_2 = -|\hat{e}| L$ at the top boundary of the domain, where $L$ is the height of the pillar in the (undeformed) initial configuration at $t=0$.
    \item The applied traction in the horizontal direction is zero on the top boundary of the domain.
\end{itemize} 
For the free-standing thin films under compression loading, the boundary conditions are as follows:
\begin{itemize}
    \item $v_1 = v_2 = 0$ at the bottom boundary of the domain.
    \item $v_1 = 0$ and $v_2 = -|\hat{e}| L$ at the top boundary of the domain. The lateral constraint $v_1 = 0 $ on the top and bottom boundaries is imposed to model the effects of resistance to material flow along the width of the film due to the ceramic blocks.
\end{itemize} 
For the free-standing thin films under simple shear loading, the boundary conditions are \cite{arora2020unification}:
\begin{itemize}
    \item $v_1 = v_2 = 0$ at the bottom boundary of the domain.
    \item $v_1 = \hat{\Gamma} L$ and $v_2 = 0$ at the top boundary of the domain. 
    \item $v_1 = \hat{\Gamma} y$ and $v_2 = 0 $ at the left and right boundaries of the domain, where $y$ is the difference between the $x_2$ coordinate of any point on the boundary with respect to the bottom boundary.
\end{itemize}
We also define the yield strength in compression $(\sigma_0)$ in terms of the yield strength in shear $(g_0)$ for different cases: $\sigma_0 = 1.14 \sqrt{3} g_0$ for the free standing film under compression and for the micropillar sandwich with $90^\circ$ orientation, while $\sigma_0 = 2 g_0$ for the micropillar sandwich with $45^\circ$ orientation.

\noindent We then define $\tau$ as the nominal reaction shear stress in the $\bfe_1$ direction on the top boundary of the free-standing film, while $\sigma$ as the nominal reaction stress in the $\bfe_2$ direction on the top boundary of the micropillar sandwich with $90^{\circ}$ film orientation. For the micropillar sandwich with $45^{\circ}$ film orientation, $\tau = 0.5 \sigma$ is the corresponding applied nominal shear stress, calculated based on stress tensor transformation in a `global' sense. 



\begin{table}[htbp]
\centering
\begin{tabular}{ |c| c c c c c c c c c c c|} 
 \hline
 \multirow{2}{4.5em}{Parameter} & $\hat{\gamma}_0$ & $m$ & $\eta$ & $b$ & $g_0$  & $g_s$  &  $\Theta_0$  & $k_0$ & $l$ & $E$ & $\nu$ \\ 
   & ($s^{-1}$) &   &  & (\AA) & (MPa) & (MPa) &  (MPa) & & ($\mu m$) & (GPa) &  \\ 
 \hline
  Value & 0.001  & 0.03 & $\frac{1}{3}$ & 4.05 & 17.3   & 161   & 392.5  & 20  & $\sqrt{3} \times 0.1$  & 62.78 & 0.3647 \\
 \hline
\end{tabular}
\caption{Material parameters for metal.}
\label{table:metal}
\end{table}

\begin{table}[htbp]
\centering
\begin{tabular}{|c| c c|} 
 \hline
 Parameter & $E$ (GPa) & $\nu$  \\ 
 \hline
Value & 110.0   & 0.20 \\
\hline
\end{tabular}
\caption{Material parameters for ceramic.}
\label{table:ceramic}
\end{table}
The parameter values used for all the simulations here are given in Table \ref{table:metal} (identical to \cite{arora2020unification} for the metal) and Table \ref{table:ceramic} for the ceramic. These are the typical material constants for copper and silicate ceramics. 
Here, $E$ and $\nu$ are the Young's modulus and the Poisson's ratio of the materials.

\subsection{Size effects} \label{size_effects_section}
\begin{figure}[H]
    \centering
    \subfloat[][Stress-strain curve]{
    \includegraphics[scale=0.35]{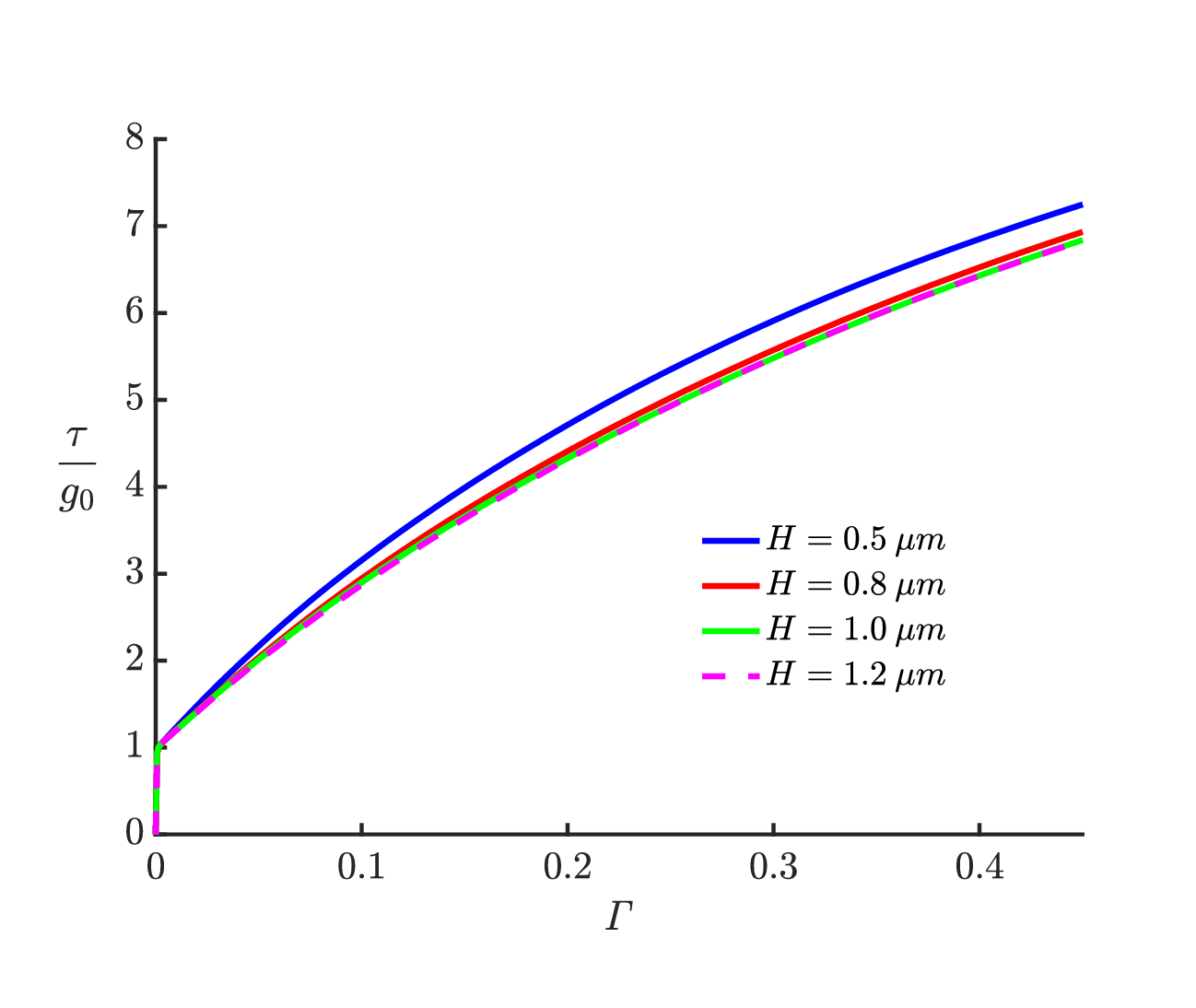}}
    \subfloat[][Power law fitted curve]{
    \includegraphics[scale=0.35]{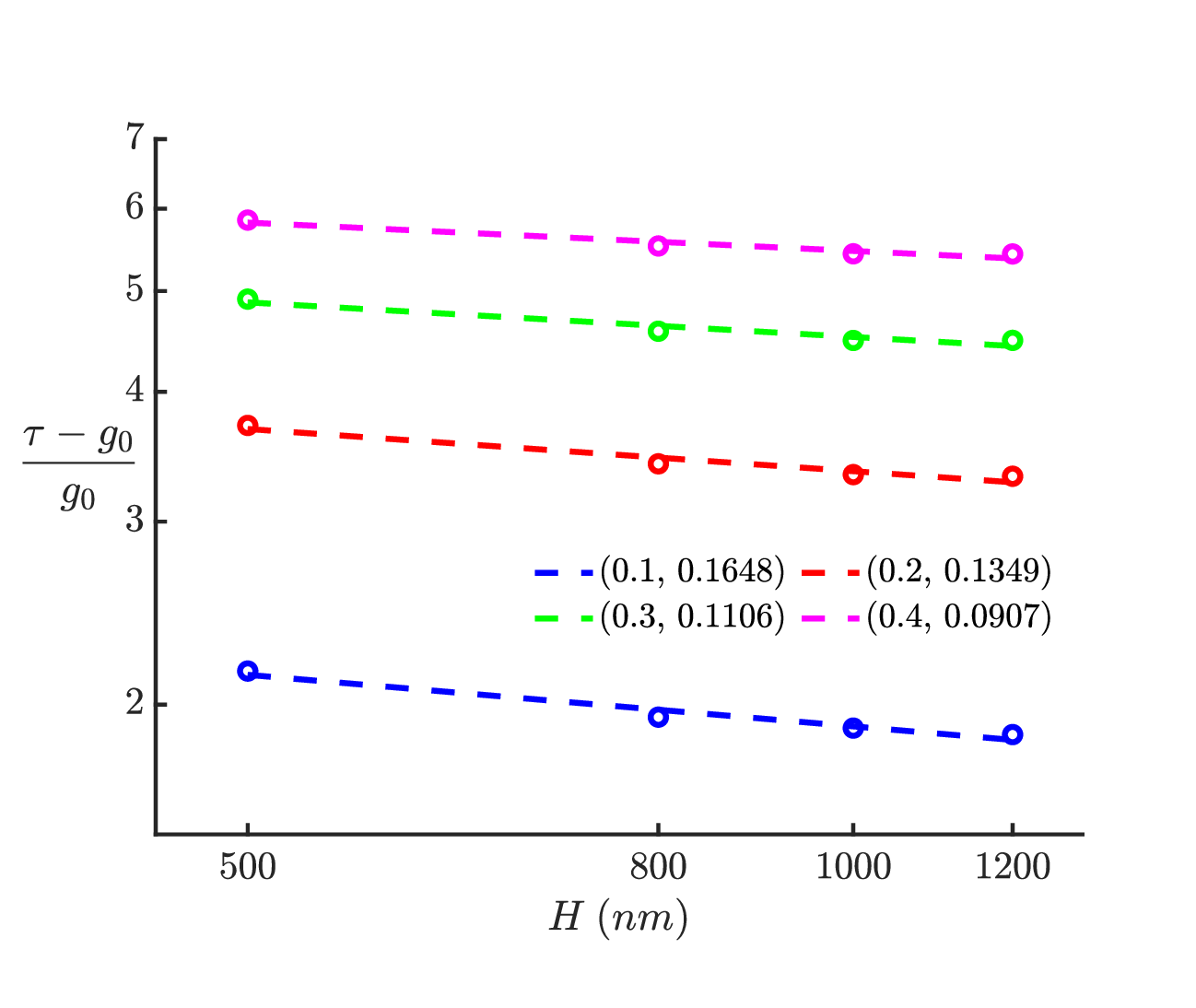}}
    \caption{(a) Stress-strain curve in simple shear of the free-standing film for different film thickness $(H)$.
    (b) Power law fitted for the stress vs film thickness ($(\tau -g_0)/g_0 = a \: H^m$) at $10\% - 40\%$ strain. The first and second term in the legend denote the strain and magnitude of power law exponent $(|m|)$, respectively.}
    \label{fig:shear_test_metal}
\end{figure}

\begin{figure}[H]
    \centering
    \subfloat[][Stress-strain curve]{
    \includegraphics[scale=0.35]{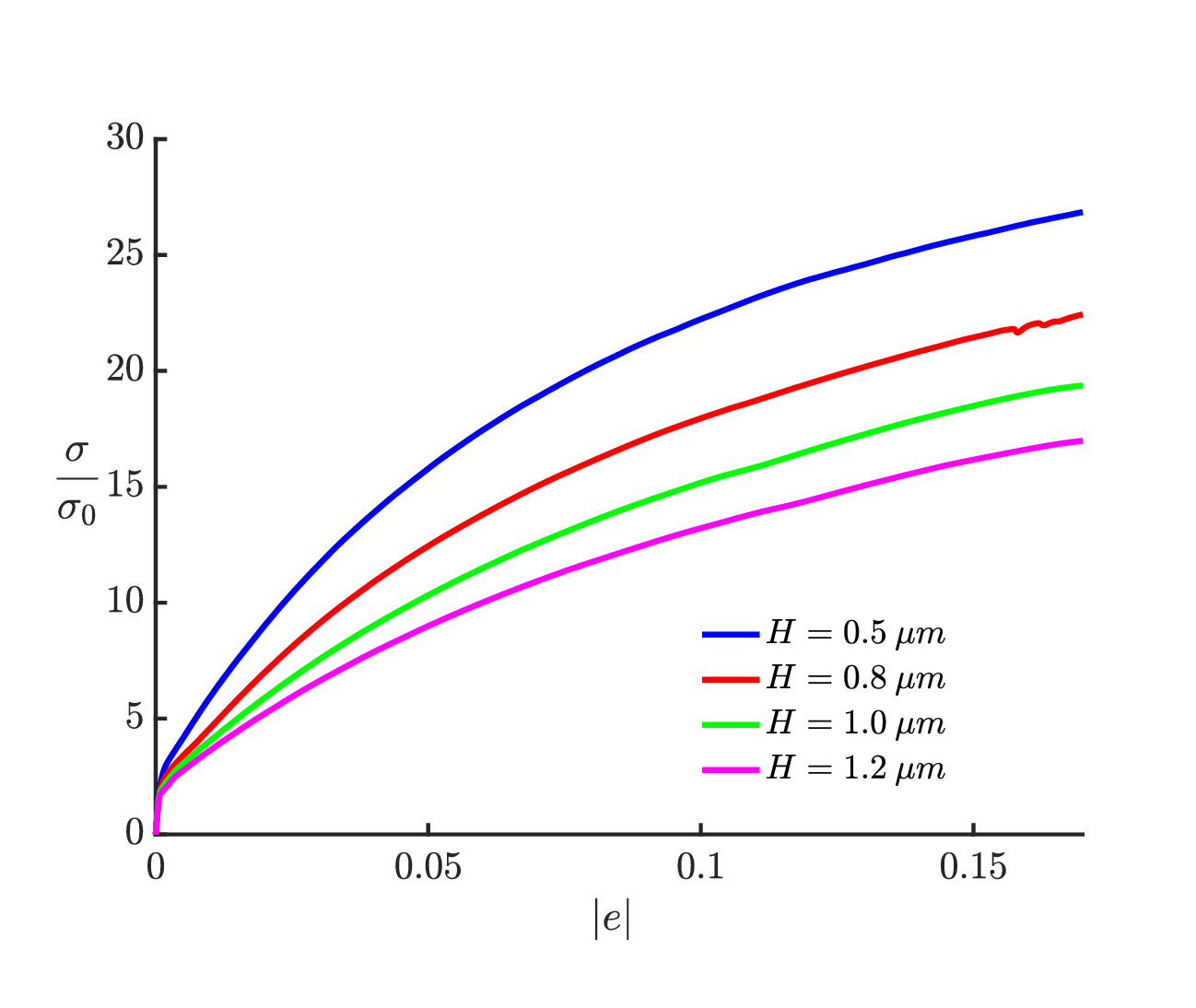}}
    \subfloat[][Power law fitted curve]{
    \includegraphics[scale=0.35]{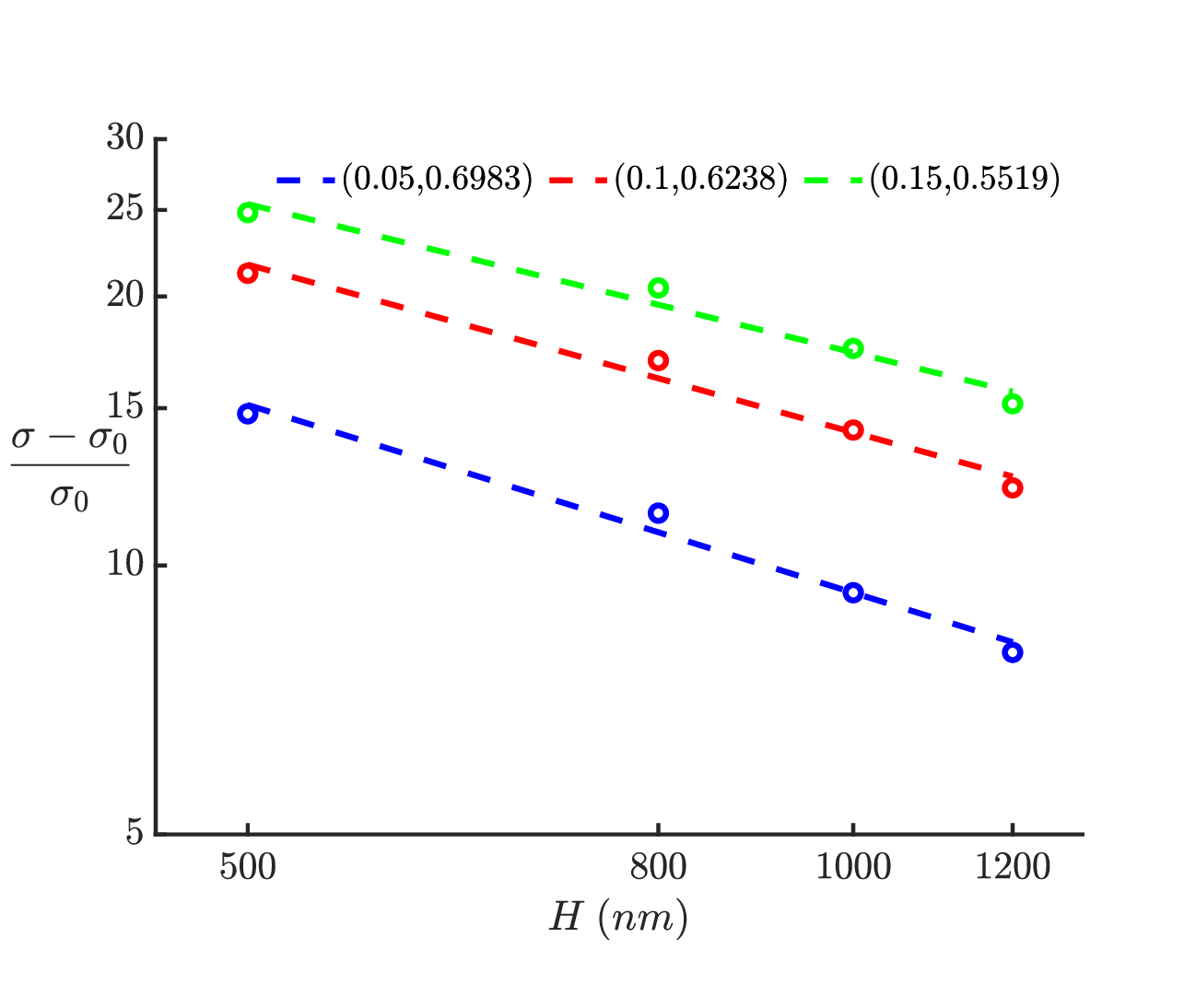}}
    \caption{(a) Stress-strain curve in compression of the free-standing film for different film thickness $(H)$.
    (b) Power law fitted for the stress vs film thickness ($(\sigma -\sigma_0)/\sigma_0 = a \: H^m$) at $5\% - 15\%$ strain. The first and second term in the legend denote the strain and magnitude of power law exponent $(|m|)$, respectively.}
    \label{fig:compression_test_metal_only}
\end{figure}

We begin with the size effect results for the free-standing thin films under simple shear and compression loadings. These simulations are an idealization of the compression experiments on sandwiched thin films in the $45^\circ$ and $90^\circ$ orientations within micropillars. These simplified cases already provide the basic explanation for the different behavior of the micropillars with thin films in the two orientations. As mentioned earlier, the constrained boundary conditions reflect the constraints of the ceramic blocks on the films. Without these constraints the deformation is essentially homogeneous and no size effect is observed. In the case of compression of free-standing films, a weak size effect is observed when no lateral constraint on material velocity is applied at the top and bottom boundaries, but with the no plastic flow condition in effect, and this is similar to the size effect in case of simple shearing (Fig.~\ref{fig:shear_test_metal}) of free-standing films. 
The shear results were also obtained in \cite[Sec.~4.1]{arora2020unification}, in accord with the experimental trends for the film in the $45^\circ$ orientation. With the lateral b.c.~constraint on material velocity in compression, a strong size effect in compression, shown in Fig.~\ref{fig:compression_test_metal_only}, is observed. As already observed in \cite{mu2016dependence}, this is essentially due to the inhomogeneous lateral material deformation, from the top and bottom boundaries to the center of the film, induced by the material velocity boundary conditions on the film; such inhomogeneity is absent in simple shear loading with the no-plastic flow constraint. In both cases, we observe that the metal film with the smallest thickness hardens the most, due to higher gradients in $\bfL^p$ for smaller domain sizes (by scaling arguments). 


An important point about the no-plastic flow b.c., related to the geometric understanding of plastic strain rate produced by the motion of dislocation lines, is that it does not restrict a simple shearing plastic flow mode in any way \cite{acharya2006size, gurtin2005boundary}. MFDM is faithful to this requirement in its b.c. implementation (whereas SGP theories known to us are not). Thus, as explained in \cite{arora2020unification}, the GND boundary layer arises at a boundary, with normal say in the $2$ direction and under $12$ simple shearing, due to the activation of $21$ components of plastic straining in $J_2$ plasticity under a $T_{12}$ shear stress. Hence, save for this feature, it would not be possible to produce inhomogeneous flow, and hence size effects, under simple shearing conditions even with the no-plastic-flow b.c. in place. Regardless, the inhomogeneity is much stronger in the case of compression with lateral b.c~constraint on material velocity, and hence the size effects in simple shear are much weaker than that in compression with lateral constraint.

In fact, in the more complex micropillar deformations to be subsequently discussed, essentially the same argument holds. In pillar compression with film in the $45^\circ$ orientation, simple shearing deformation conditions are realized on the film on average even with a no-plastic-flow condition imposed due to the presence of ceramic layers, and a much weaker size effect is observed, in comparison to pillar compression with film in the $90^\circ$ orientation where the lateral constraint on material deformation is imposed due to the presence of the ceramic blocks.

With this basic understanding of the gross behavior, we then model the compression of the various micropillar configurations. Interestingly, we again recover the size-effect trends observed in experiment, but we demonstrate significant differences in local mechanical fields, in comparison to the free-standing films, that we describe in Sec.~\ref{local_fields}.

\begin{figure}[H]
    \centering
    \subfloat[][Stress-strain curve]{
    \includegraphics[scale=0.35]{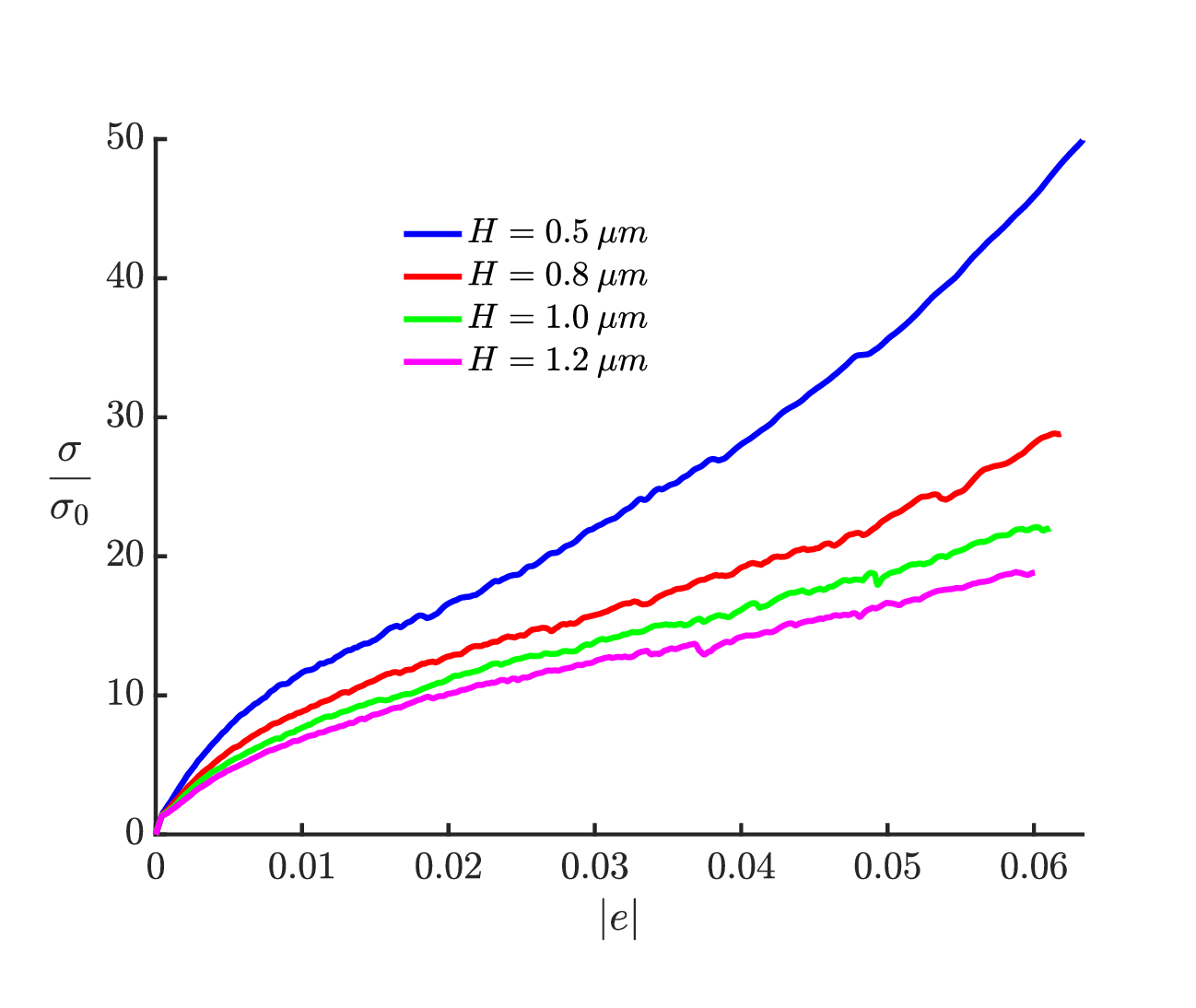}}
    \subfloat[][Power law fitted curve]{
    \includegraphics[scale=0.35]{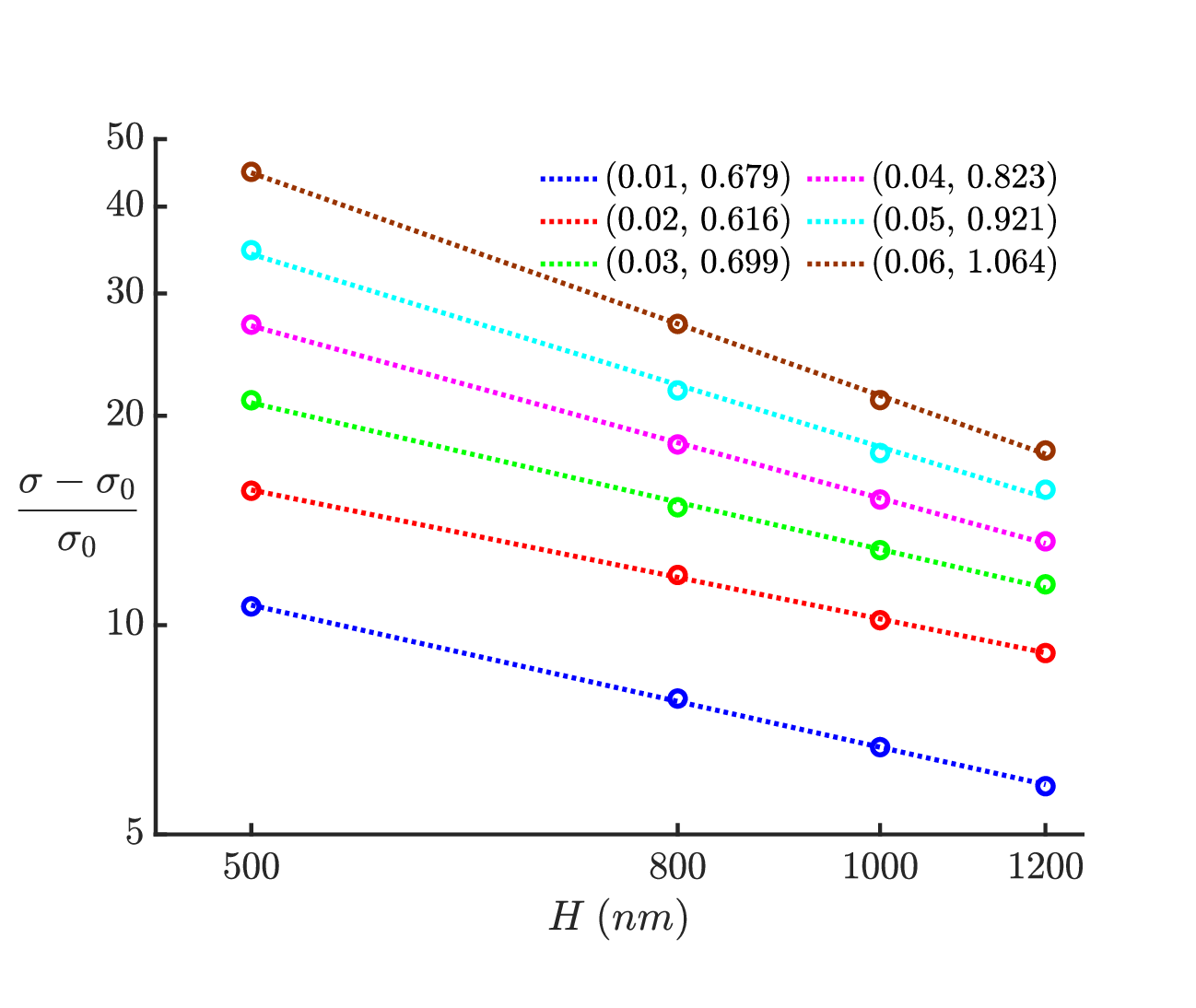}}
    \caption{(a) Stress-strain curve in compression of the micropillar sandwich with $90^{\circ}$ film orientation and for different film thickness $(H)$ .
    (b) Power law fitted for the stress vs film thickness ($(\sigma -\sigma_0)/\sigma_0 = a \: H^m$) at $1\% - 6\%$ strain. The first and second term in the legend denote the strain and magnitude of power law exponent $(|m|)$, respectively.}
    \label{fig:compression_test_90}
\end{figure}

\begin{figure}[htbp]
\centering
\subfloat[][$H = 0.5 \: \mu m$]{
\includegraphics[scale=0.85]{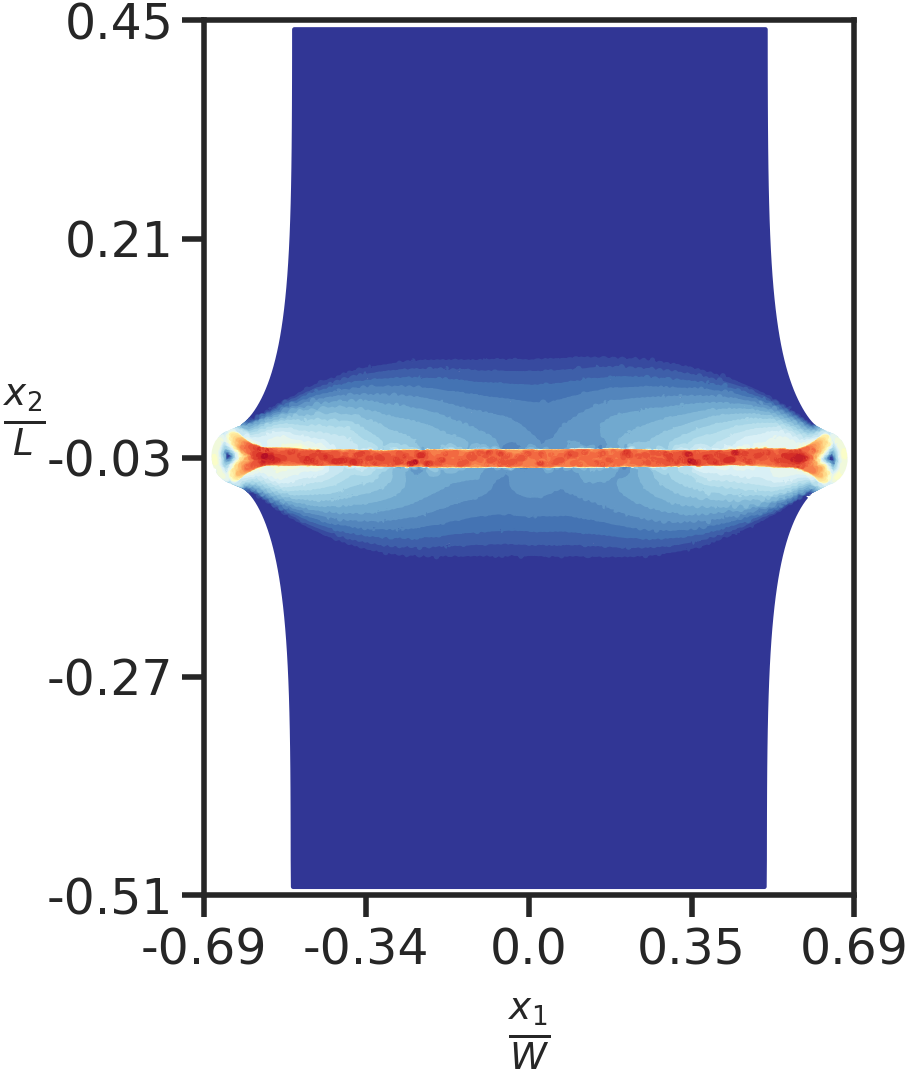}}
\subfloat[][{Deformed shape under compression}]{
\includegraphics[scale=0.53]{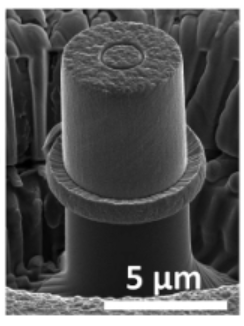}}
\subfloat[][$H = 1.2 \: \mu m$]{
\includegraphics[scale=0.85]{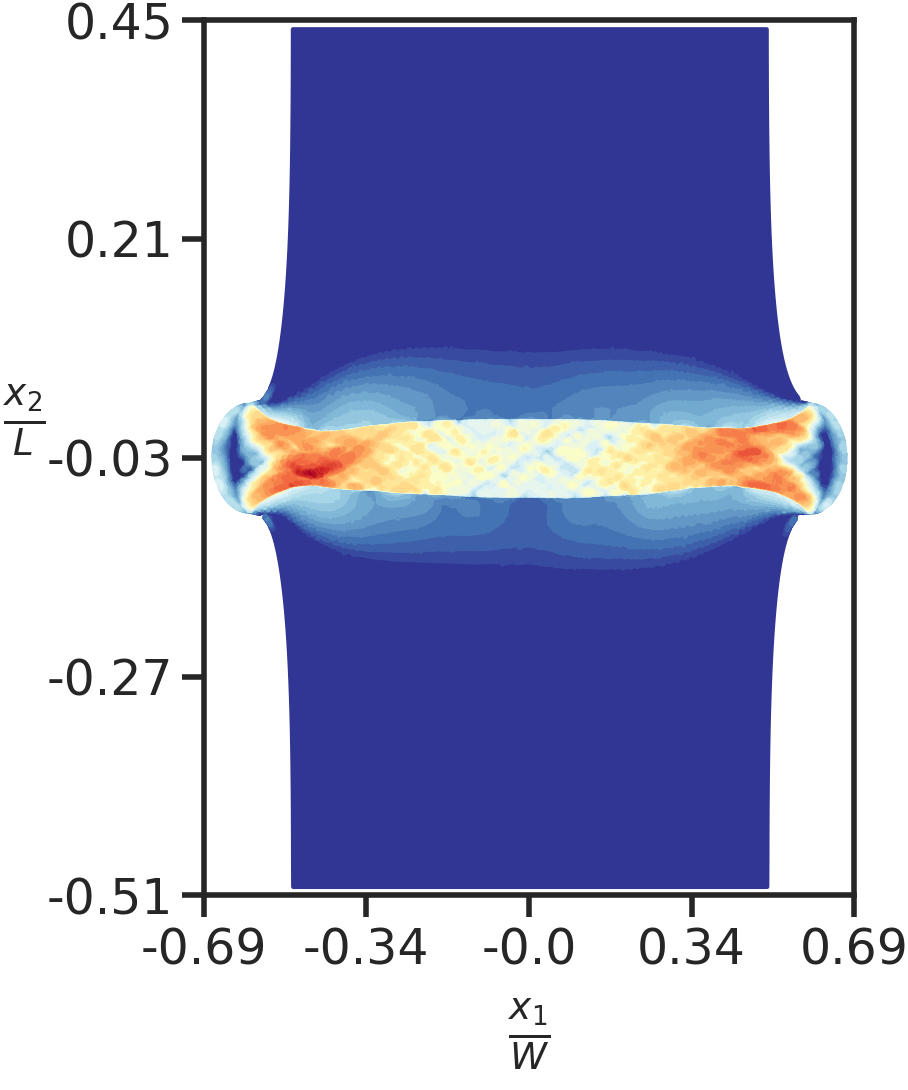}}
\quad
\subfloat{
\includegraphics[width = 0.45\textwidth]{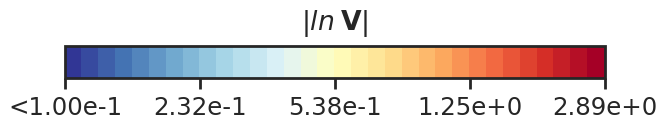}}
\caption{Norm of Logarithmic strain tensor $(|ln (\bs{V})|)$ at 6\% compression strain $(|e|)$ for the micropillar sandwich with $90^{\circ}$ film orientation for different film thickness $(H)$, and comparison of the deformed shape with the experimental results \cite{zhang2017mechanical} . (Figure in (b) reprinted from \cite{zhang2017mechanical} with permission from \textit{Elsevier}).}
\label{fig:fig_rho_g_90_full}
\end{figure}

Fig.~\ref{fig:compression_test_90} shows the stress-strain curves and the fitted power law relationship between the stress and film thickness i.e., $(\sigma -\sigma_0)/\sigma_0 = a \: H^m$. Here, $|m|$ approaches 1.0 at $6\%$ overall strain, and the same value for the exponent is reported in the experimental work of Meng and co-workers \cite{mu2016dependence} on Cu-CrN as-deposited micropillars, shown in Fig.~\ref{fig:experiment}(a). As shown in Fig.~\ref{fig:fig_rho_g_90_full}, for the micropillar sandwich with the thin film in the $90^{\circ}$ orientation, the film keeps bulging out with increase in the applied compressive strain, as similarly observed in the post test scanning electron microscope images of Cu interlayers in \cite{zhang2017mechanical}. Fig.~\ref{fig:fig_rho_g_90_full} also shows the norm of the Logarithmic strain tensor, $|ln \: (\bs{V})|$ at $6 \%$ overall compression strain for different film thickness, where $\bs{V}$ is the left stretch tensor of the Polar Decomposition of the deformation gradient $\bfF$ from the initial configuration at $t=0$.

\begin{figure}[H]
    \centering
    \subfloat[][Stress-strain curve]{
    \includegraphics[scale=0.35]{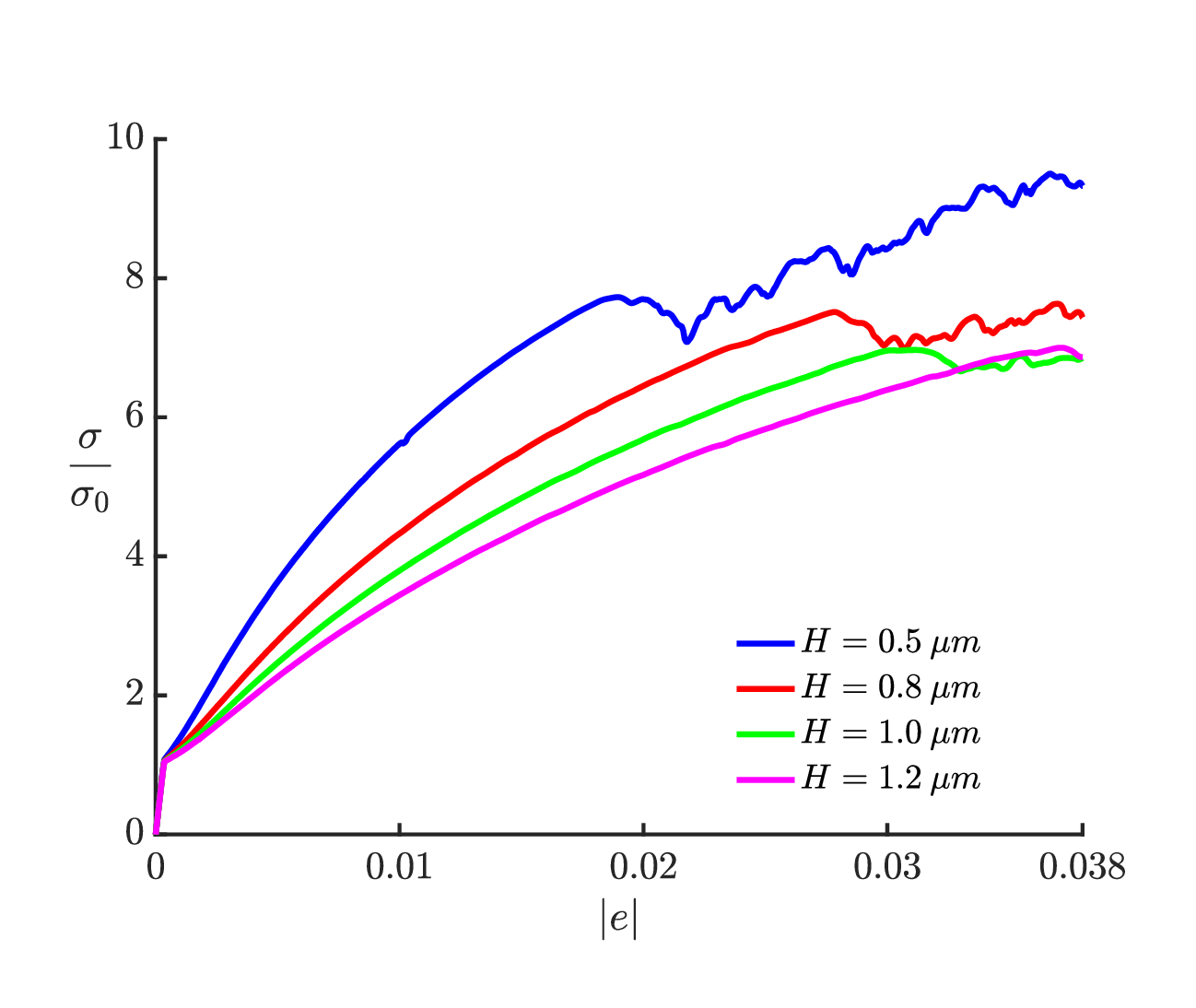}}
    \subfloat[][Power law fitted curve]{
    \includegraphics[scale=0.35]{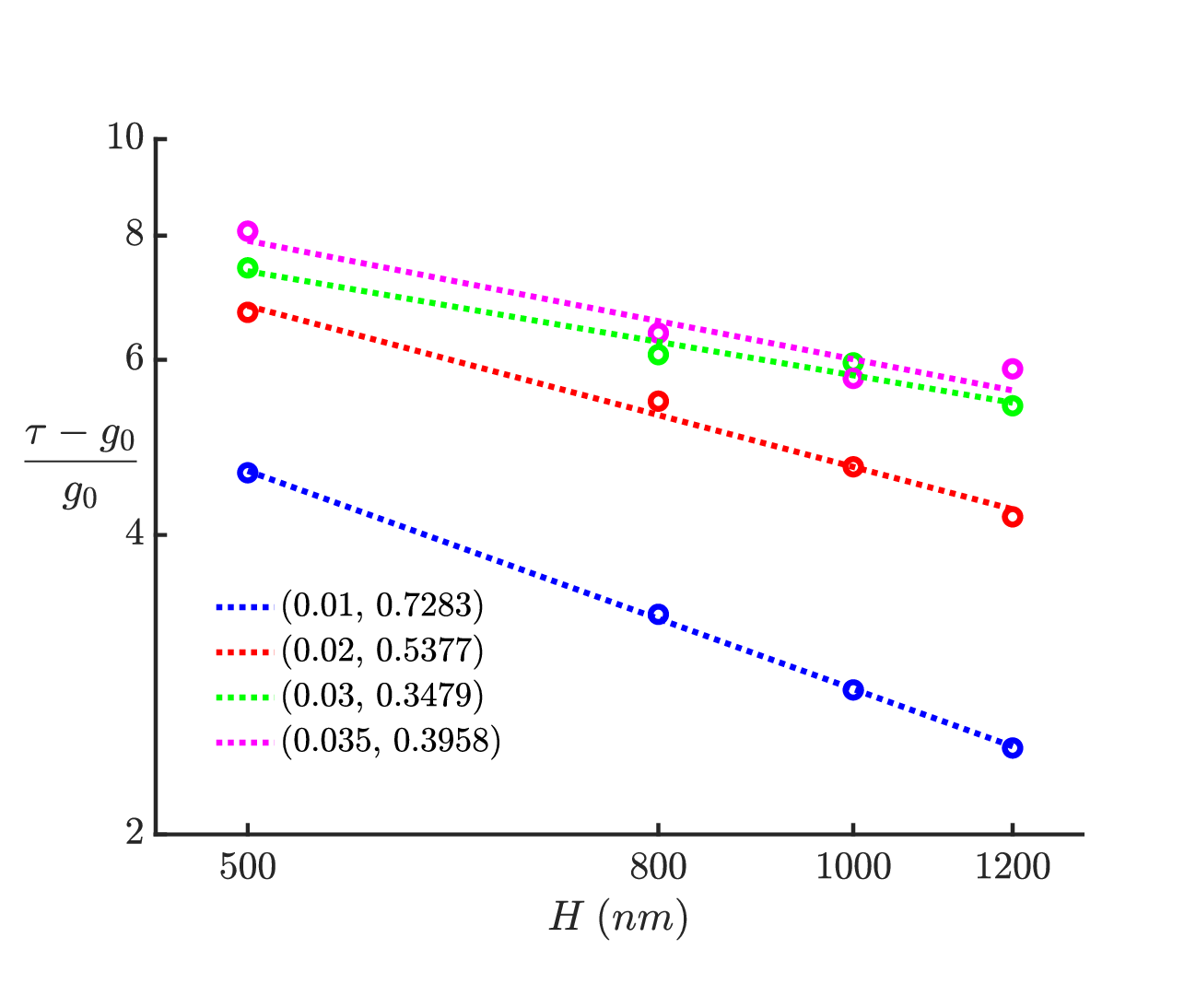}}
    \caption{(a) Stress-strain response in compression of the micropillar sandwich with $45^{\circ}$ orientation and for different film thickness $(H)$.
    (b) Power law fitted for the stress vs film thickness ($(\tau -g_0)/g_0 = a \: H^m$) at $1\% - 3.5\%$ strain. The first and second term in the legend denote the strain and magnitude of power law exponent $(|m|)$, respectively.}
    \label{fig:compression_test_45}
\end{figure}

Fig.~\ref{fig:compression_test_45}(a) shows the stress-strain response for micropillars with the metal thin film in the $45^\circ$ orientation, and  Fig.~\ref{fig:compression_test_45}(b) shows the fitted power-law relationship between the applied nominal shear stress $(\tau = 0.5 \sigma)$ 
and film thickness i.e., $(\tau - g_0)/g_0 = a \: H^m$. The magnitude of the power-law exponent $(|m|)$ decreases from 0.728 to 0.347 as $|e|$ increases from $1.0\%$ to $3.0\%$. The $|m|$ obtained at $3.5\%$ strain is 0.395, however, if only samples with $H >= 0.8 \: \mu m$ are considered, then $|m|$ obtained will be close to zero.  The simulations were stopped as the metal film region was close to penetration into the elastic blocks as shown later in Fig.~\ref{fig:45_degree_0_5_before_penetration}. The magnitude of the power-law exponent reported in \cite{mu2016dependence} for the $45^{\circ}$ orientation pillar with as-deposited samples is 0.2, as shown in Fig.~\ref{fig:experiment}(b). Upon further straining, if possible without penetration, the value of power-law exponent ($m$) will reduce. This can be justified based on the reducing trend in the value of $m$ observed for the simple shearing of free-standing films. The value of $m$ reduces to $0.0907$ at $40\%$ strain (Fig.~\ref{fig:shear_test_metal} of this paper). Similar trends for the value of $m$ were also obtained with increase in strain in \cite[Fig.~3]{arora2020unification}. 

Our model for the plastic straining due to statistical dislocations (i.e.~$\bfL^p$) is phenomenological and it is expected that as the overall length scale over which plastic flow occurs decreases, the contribution of $\bfL^p$ in the total plastic strain rate for the model should decrease (physically, there are fewer and fewer sources, but the phenomenological $J_2$  plasticity model assumes that there is an abundant supply of sources, and all that is required is stress to mobilize them). This is not accounted for in the current model and at small scales, e.g.~$H=0.5 \: \mu m$, there is excessive hardening due to higher gradients in $\bfL^p$.

An eventual reduction in the hardening modulus in the shear stress-strain response in Fig.~\ref{fig:compression_test_45}(a) is observed, approaching a `flat' response on average, but not the shear stress plateau as experimentally observed by \cite{mu2016dependence}. Our model does not have a failure model for the interfacial regions which would be capable of demonstrating such response under the intense local shearing observed in the experiments and in the simulations.  Kuroda et al.~\cite{kuroda2021constraint} are able to demonstrate the observed plateau for the film in the $45^\circ$ orientation in the micropillar; a small value of the hardening modulus is employed in their calculations, a material description that would be hard-pressed to reproduce the stress-strain behaviour of polycrystalline copper for macroscopic pillar and film sizes. 

For brevity, the comparisons of the size effects obtained from our simulations with those obtained in the experimental work of \cite{mu2016dependence} are shown in Table \ref{tab:comparison_of_size_effects}.

\begin{table}[H]
    \centering
    \begin{tabular}{|c |c| c|} \hline
       Magnitude of the & Our simulations  & \multirow{2}{*}{Experiment \cite{mu2016dependence}} \\ 
       power law exponent & (before self-contact) & \\
       \hline
        Micropillar $45^{\circ}$ orientation & 0.395 & \multirow{2}{*}{0.2}\\ 
        Free-standing film (shearing) & 0.0907 &  \\ \hline
        Micropillar $90^{\circ}$ orientation & 1.0 & \multirow{2}{*}{1.0} \\
        Free-standing film (compression) & 0.5519 & \\
         \hline
    \end{tabular}
    \caption{Magnitude of power law exponent in size effect results from our simulations (before self-contact of metal films into elastic blocks, particularly, for micropillar with $45^{\circ}$ orientation) and experiments \cite{mu2016dependence}.}
    \label{tab:comparison_of_size_effects}
\end{table}

\subsection{Local fields in the free-standing film vs. the micropillar sandwich}\label{local_fields}
In this section, we show the significant differences in the local mechanical response of the free-standing films and the micropillar-film sandwich (which, nevertheless, produce the same qualitative size effects, as already shown).

\textit{When comparing the local fields for the micropillar sandwich with the film in the $45^{\circ}$ orientation and the free-standing film, the film for the micropillar case is rotated by $45^{\circ}$ in an anti-clockwise sense, for the ease of visualization}.

In the following, we define $|e_m|$ (the subscript $m$ stands for metal) as the nominal compressive strain in the metal film corresponding to a given nominal compressive strain ($|e|$) for the entire domain - for the free-standing film, $|e_m| = |e|$.  As discussed earlier, for the micropillar sandwich, $|e|$ is the magnitude of the engineering compressive strain calculated from the applied boundary condition on the top of the pillar and the initial length of the pillar.

We also define $\Gamma_m$ as the nominal shear strain in the metal film for the micropillar sandwich as well as the free-standing film.
\subsubsection{Micropillar with thin film in the $90^{\circ}$ orientation}

\begin{figure}[H]
\centering
\subfloat[][$H = 0.5 \: \mu m$ (free-standing film)]{
\includegraphics[width=0.45\textwidth]{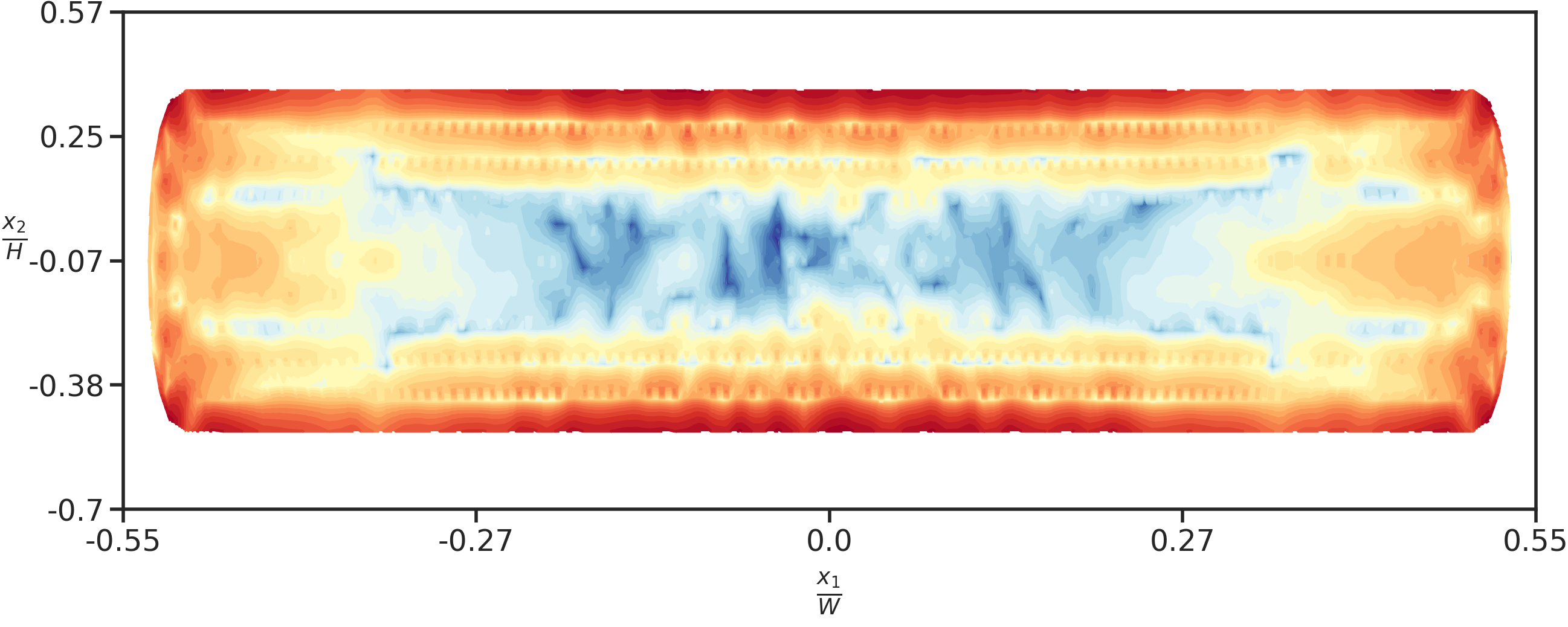}}
\subfloat[][$H = 0.5 \: \mu m$ (micropillar)]{
\includegraphics[width=0.45\textwidth]{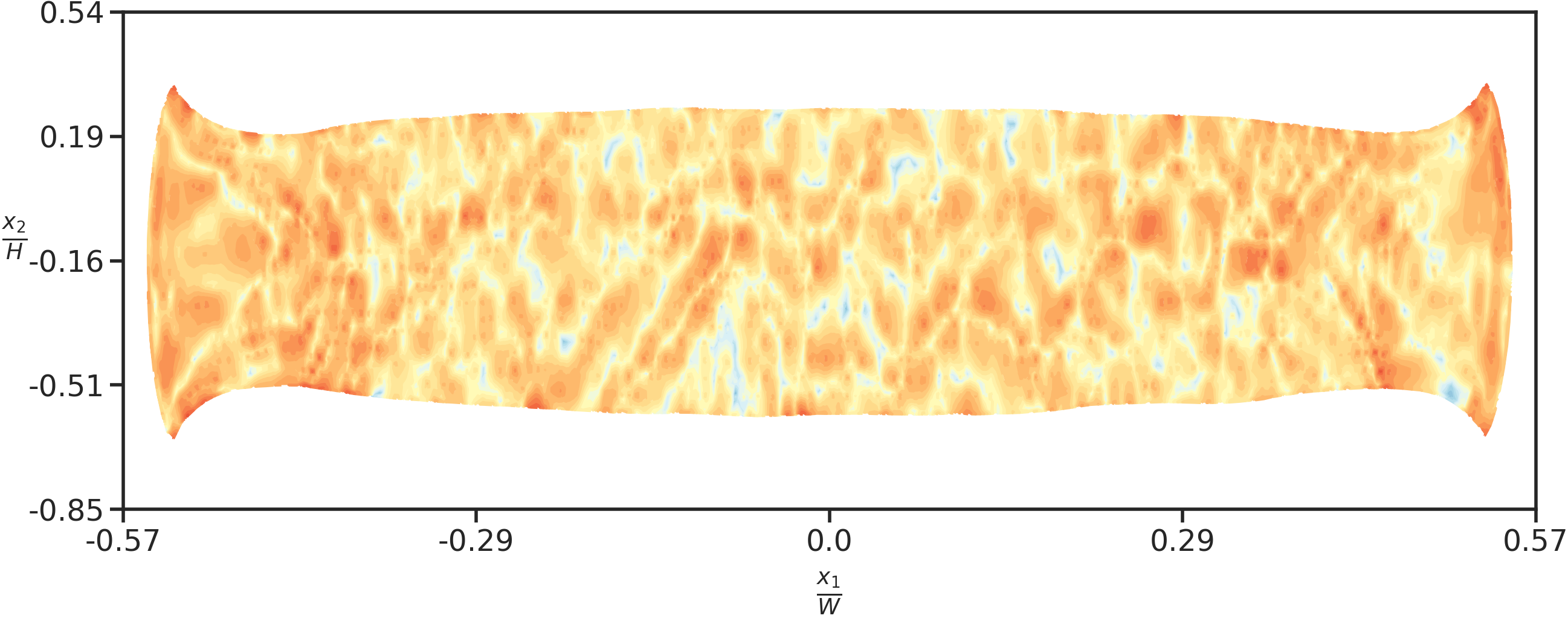}}
\quad
\subfloat{
\includegraphics[width = 0.45\textwidth]{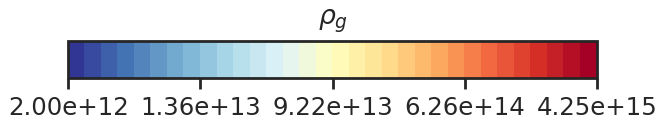}}
\caption{$\rho_g  = |\bs{\alpha}|/b \: (m^{-2})$ at $|e_m| = 15\%$ in the metal film for the free-standing film and the micropillar sandwich with $90^{\circ}$ orientation. The corresponding $|e|$ for the micropillar is $1.56\%$.}
\label{fig:fig_rho_g_90}
\end{figure}

Fig.~\ref{fig:fig_rho_g_90} compares the $\rho_g := |\bs{\alpha}|/b$ (GND) field plots at $|e_m| = 15\%$ compressive strain in the metal film for the free-standing film and the micropillar sandwich. Excess dislocation boundary layers are absent in the micropillar film configuration. They arise in the free-standing films due to no-flow boundary conditions. The $\rho_g$ patterns observed in the micropillar film are driven by the deformation inhomogeneity induced by the the resistance to horizontal material flow at the boundaries of the film with the ceramic blocks.

\begin{figure}[H]
\centering
\subfloat[][$H = 0.5 \: \mu m$ (free-standing film)]{
\includegraphics[width=0.45\textwidth]{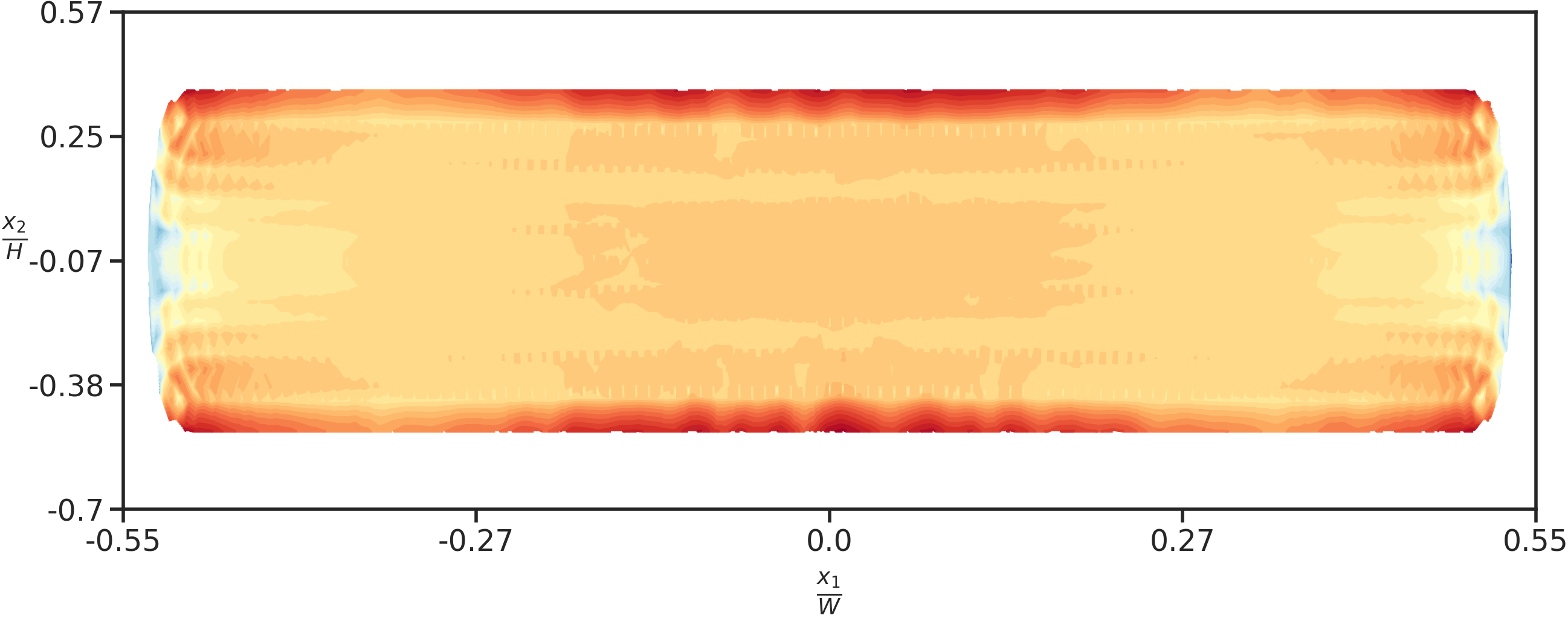}}
\subfloat[][$H = 0.5 \: \mu m$ (micropillar)]{
\includegraphics[width=0.45\textwidth]{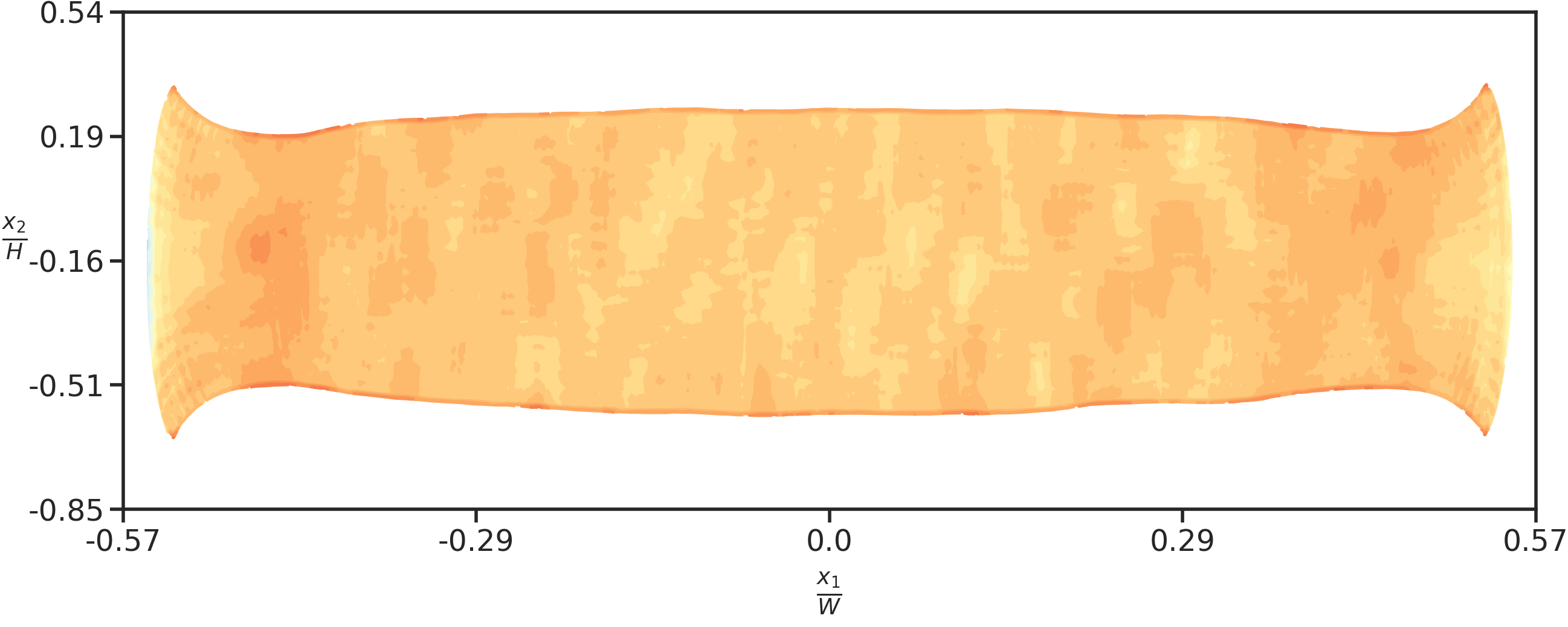}}
\quad
\subfloat{
\includegraphics[width = 0.45\textwidth]{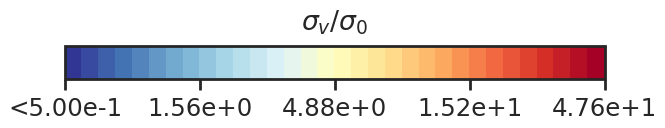}}
\caption{von Mises stress at $|e_m| = 15\%$ in the metal film for the free-standing film and the micropillar sandwich with $90^{\circ}$ orientation. The corresponding $|e|$ for the micropillar is $1.56\%$.}
\label{fig:fig_von_mises_90}
\end{figure}

\begin{figure}[H]
\centering
\subfloat[][$H = 0.5 \: \mu m$ (free-standing film)]{
\includegraphics[width=0.45\textwidth]{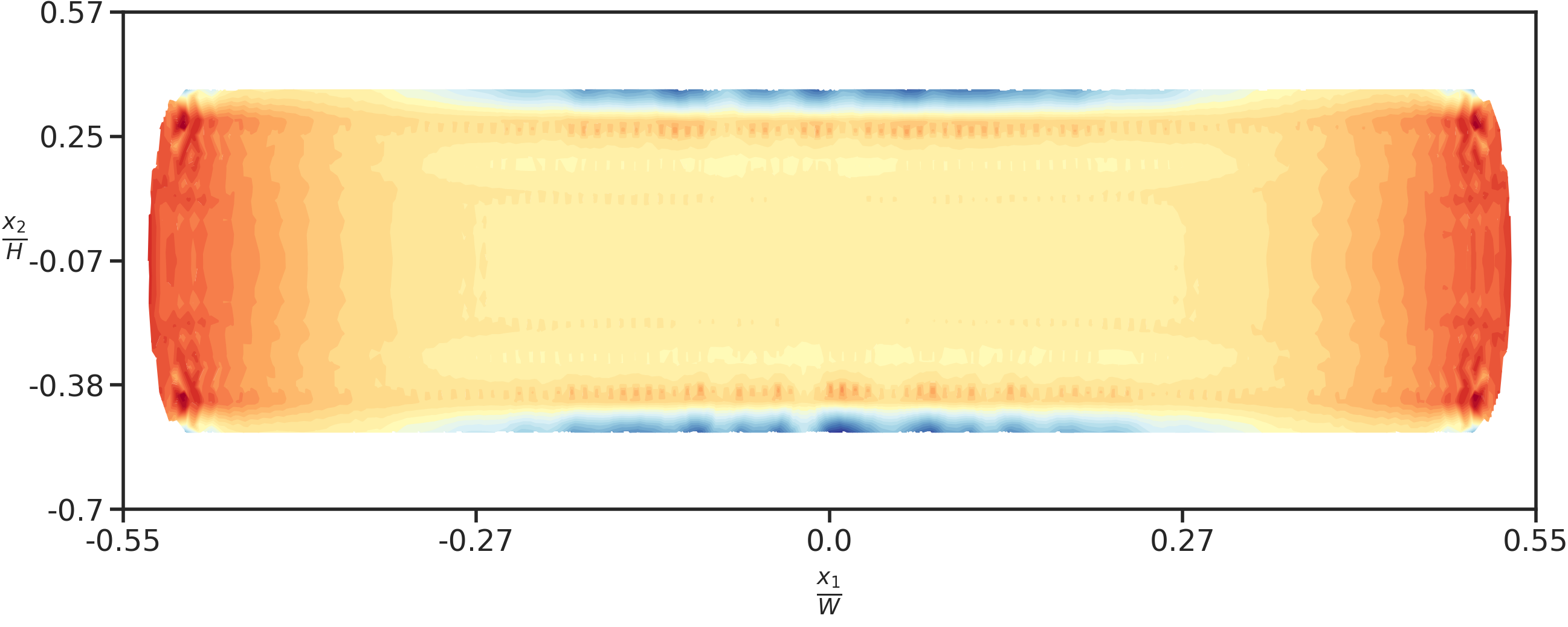}}
\subfloat[][$H = 0.5 \: \mu m$ (micropillar)]{
\includegraphics[width=0.45\textwidth]{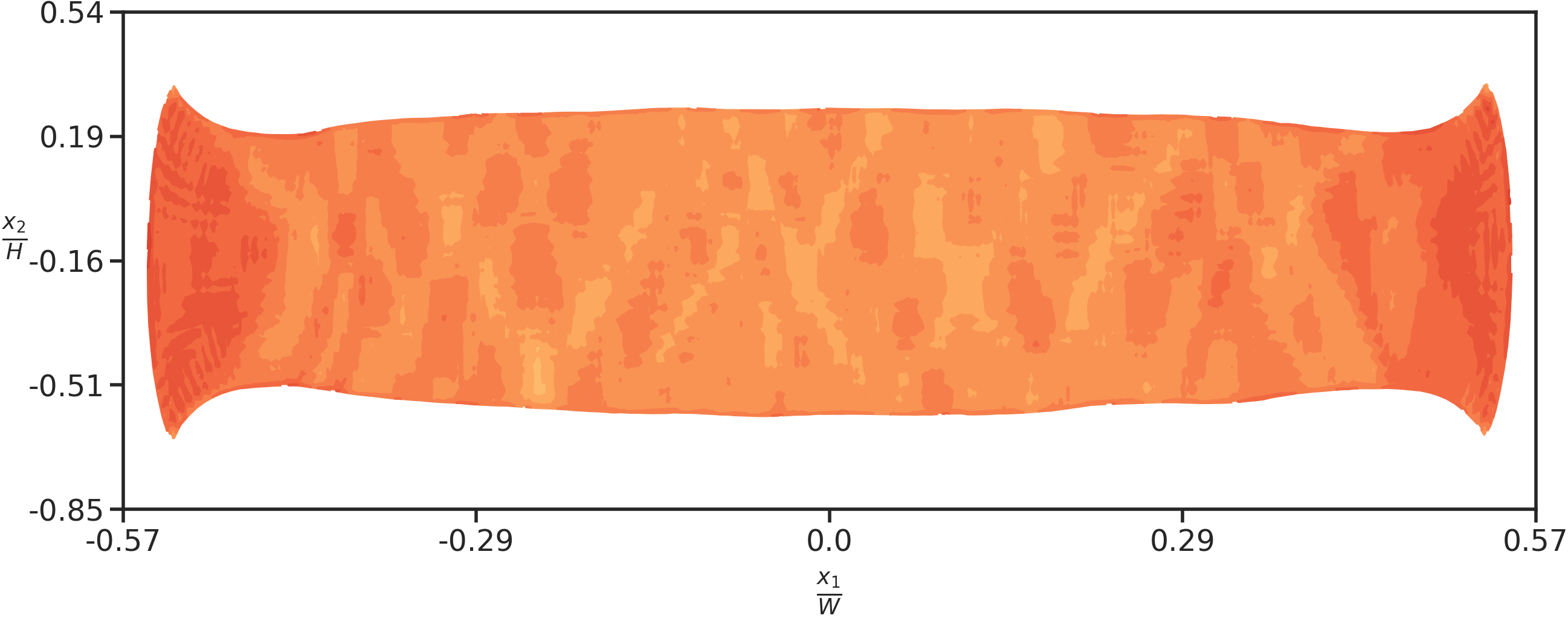}}
\quad
\subfloat{
\includegraphics[width = 0.45\textwidth]{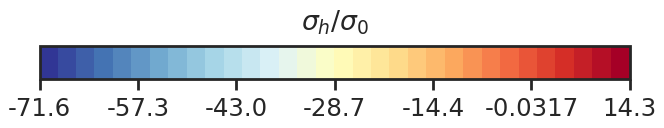}}
\caption{Hydrostatic stress at $|e_m| = 15\%$ in the metal film for the free-standing film and the micropillar sandwich with $90^{\circ}$ orientation. The corresponding $|e|$ for the micropillar is $1.56\%$.}
\label{fig:fig_pressure_90}
\end{figure}

Fig.~\ref{fig:fig_von_mises_90} shows the von Mises stress and Fig.~\ref{fig:fig_pressure_90} the hydrostatic stress fields (both normalised with yield stress) at $|e_m| = 15\%$ in the metal film for the free-standing film and the micropillar sandwich.  The magnitude of $\sigma_v$ for the free-standing film in Fig.~\ref{fig:fig_von_mises_90} is higher at the top/bottom boundary of the thin film. The hydrostatic stress is higher in the case of the free-standing film as compared to the micropillar case. The constraint to lateral motion at the top and bottom boundaries of the film is softer for the film in the micropillar sandwich, which results in the generation of less hydrostatic stress.

\subsubsection{Micropillar with thin film in the $45^{\circ}$ orientation}

\begin{figure}[H]
\centering
\subfloat[][$\Gamma_m = 1.02$ for $H = 0.5 \: \mu m$]{
\includegraphics[width=0.48\textwidth]{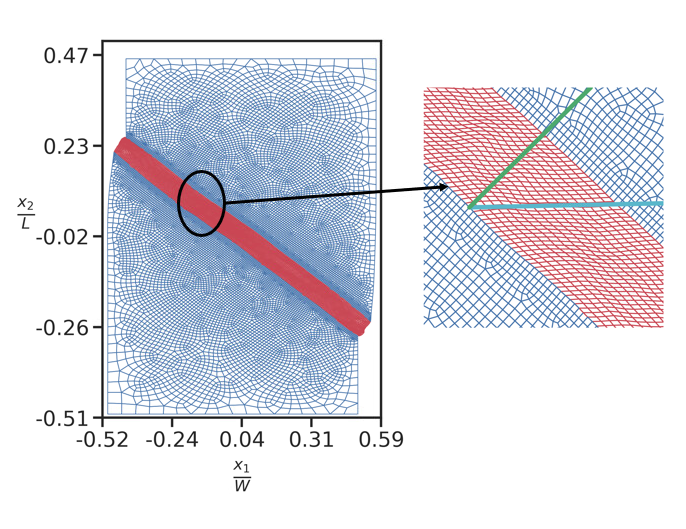}}
\subfloat[][$\Gamma_m = 0.39$ for $H = 1.2 \: \mu m$]{
\includegraphics[width=0.48\textwidth]{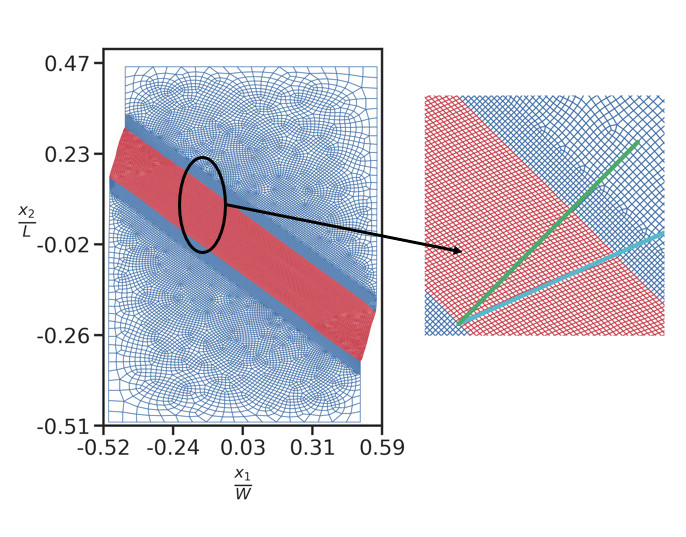}}
\caption{Nominal shear strain ($\Gamma_m$) in the film at $|e| = 3.5\%$ for micropillar with $45^{\circ}$ orientation, for different film thickness $(H)$. The green solid line denotes an undeformed line, while the cyan solid line denotes the corresponding deformed line.}
\label{fig:fig_shear_strain_45}
\end{figure}

\begin{figure}[H]
    \centering
    \includegraphics[scale=0.4]{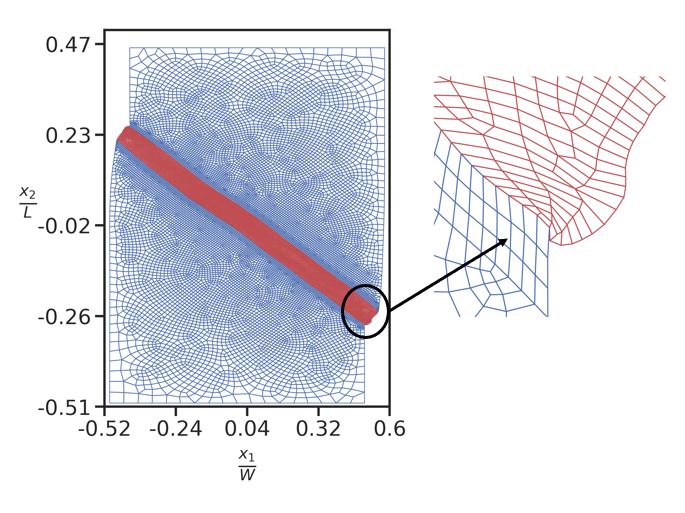}
    \caption{The deformed mesh for the micropillar with $45^{\circ}$ orientation for $H = 0.5 \: \mu m$ at $|e| = 3.8\%$, when the film material is close to penetrating into ceramic blocks.}
    \label{fig:45_degree_0_5_before_penetration}
\end{figure}

Fig.~\ref{fig:fig_shear_strain_45} shows the deformed mesh and the nominal shear strain ($\Gamma_m$) in the metal film at $|e| = 3.5\%$ compressive strain, for different film thickness. Under increased compression, the film material near the lateral boundaries of the micropillar rotate excessively due to the shearing and (damage-free) constraint of the ceramic blocks. The calculations are stopped when the film material  is close to penetrating the ceramic blocks, as shown in Fig.~\ref{fig:45_degree_0_5_before_penetration}. Although the magnitude of $|e|$ is not that high, $\Gamma_m$ in the metal film is very large. For instance, at $|e| = 3.5\%$, $\Gamma_m$ in the metal film is $102\%$ for $H=0.5\: \mu m$, while it is $39\%$ for $H=1.2\: \mu m$.

\begin{figure}[htbp]
\centering
\subfloat[][$H = 0.5 \: \mu m$]{
\includegraphics[width=0.45\textwidth]{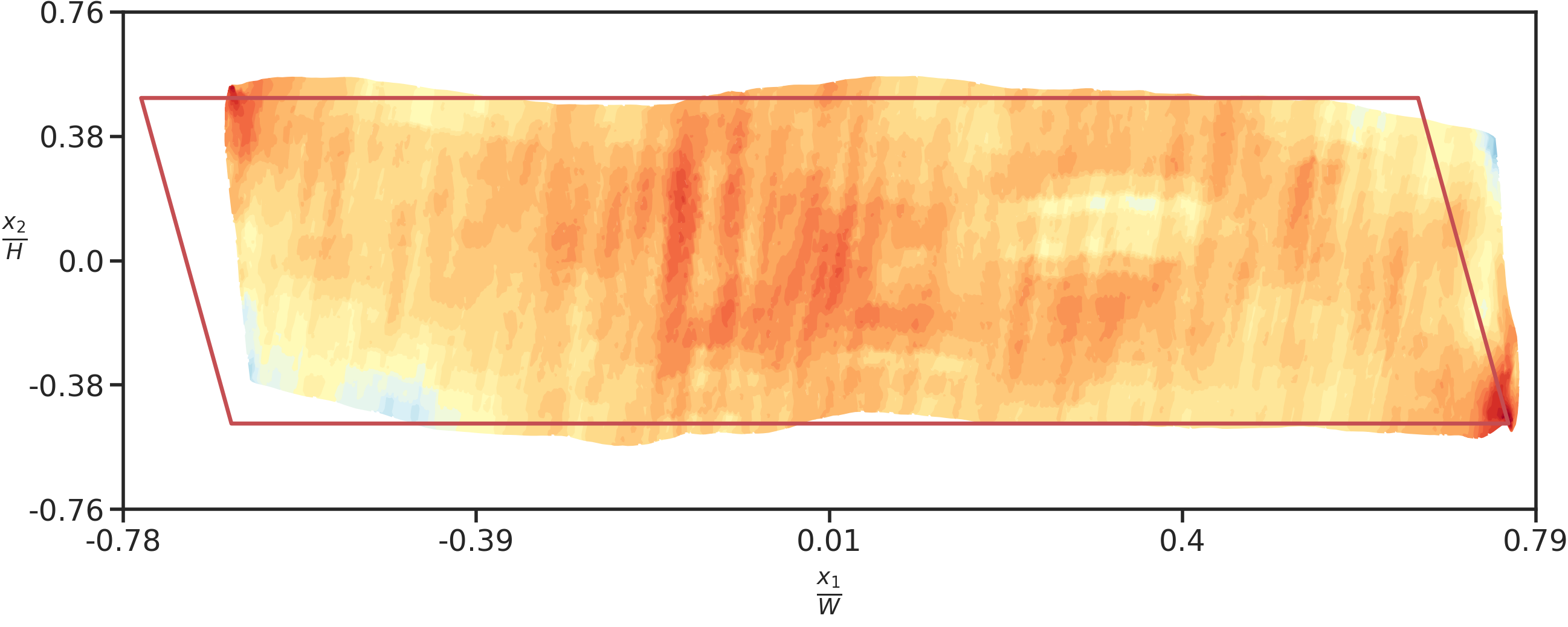}}
\subfloat[][$H = 1.2 \: \mu m$]{
\includegraphics[width=0.45\textwidth]{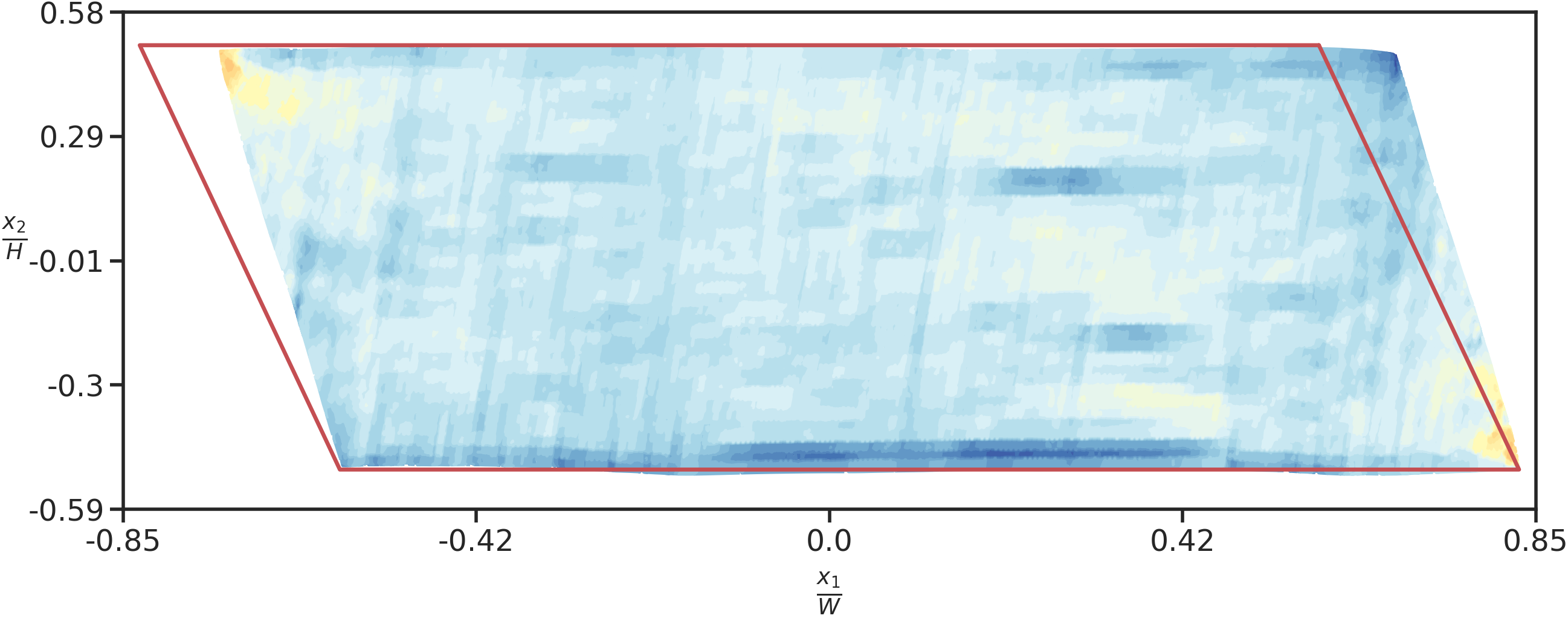}}
\quad
\subfloat{
\includegraphics[width = 0.45\textwidth]{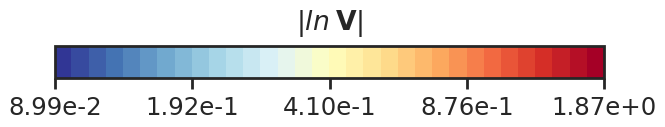}}
\caption{Norm of Logarithmic strain tensor $(|ln (\bs{V})|)$ at 3.5\% compression strain in the micropillar sandwich with $45^{\circ}$ orientation, for different film thickness $(H)$. The solid red line shows the boundary of the undeformed metal thin film.}
\label{fig:fig_rho_g_45_full}
\end{figure}

\begin{figure}[htbp]
\centering
\subfloat[][$H = 0.5 \: \mu m$ (micropillar)]{
\includegraphics[width=0.45\textwidth]{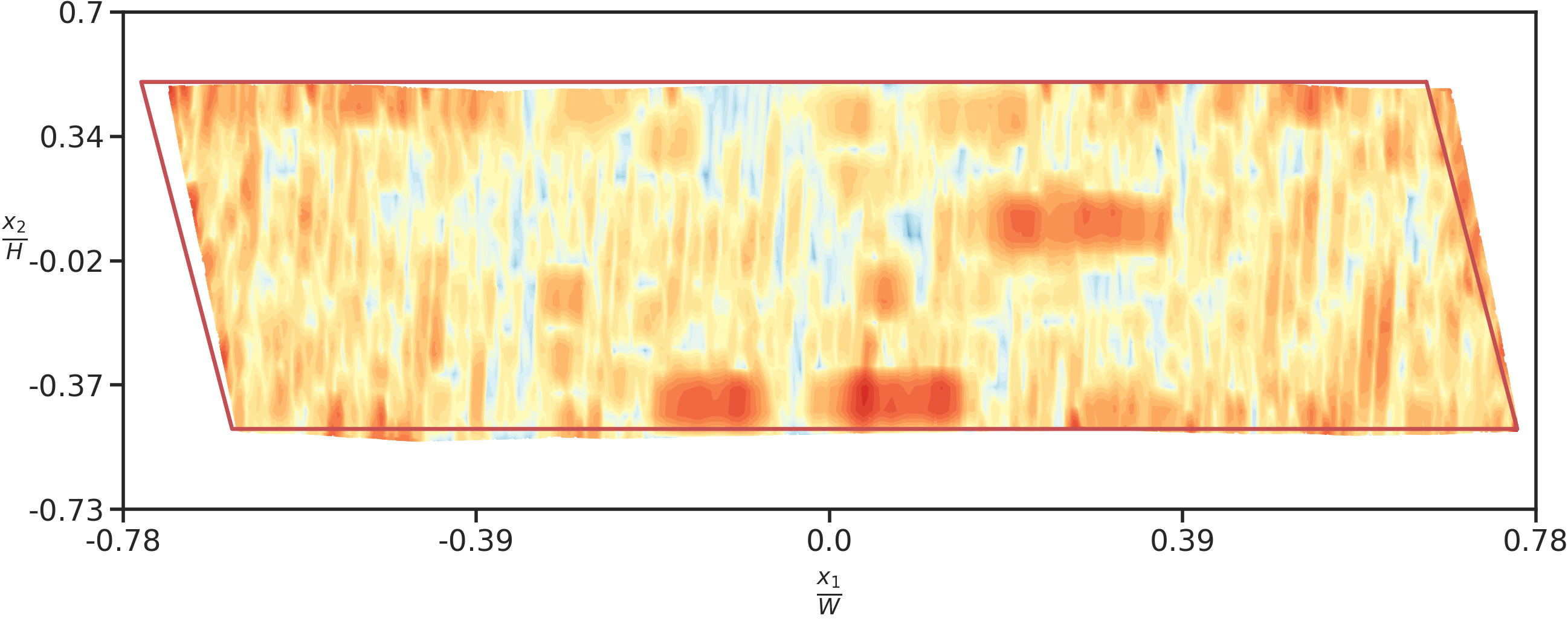}}
\subfloat[][$H = 0.5 \: \mu m$ (free-standing film)]{
\includegraphics[width=0.45\textwidth]{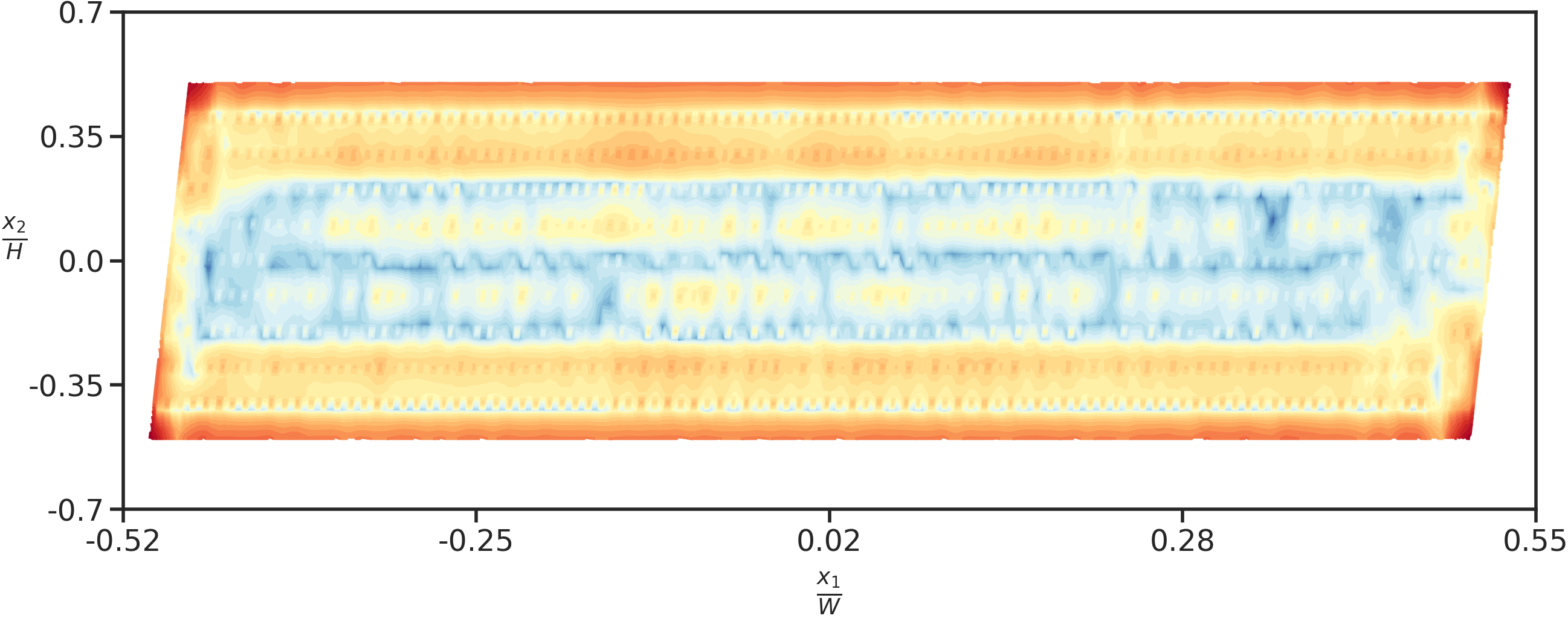}}
\quad
\subfloat{
\includegraphics[width = 0.45\textwidth]{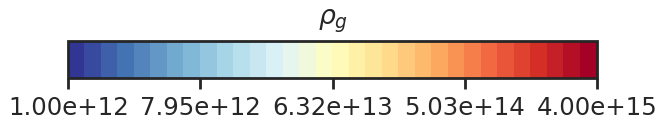}}
\caption{$\rho_g  = |\bs{\alpha}|/b \: (m^{-2})$ at $\Gamma_m = 30\%$ in the metal film for the micropillar sandwich with $45^{\circ}$ orientation and for the free-standing film. The corresponding $|e|$ for the micropillar is $1.15\%$, and the solid red line in (a) shows the boundary of the undeformed metal thin film.}
\label{fig:fig_rho_g_45}
\end{figure}

\begin{figure}[htbp]
\centering
\subfloat[][$H = 0.5 \: \mu m$ (micropillar)]{
\includegraphics[width=0.45\textwidth]{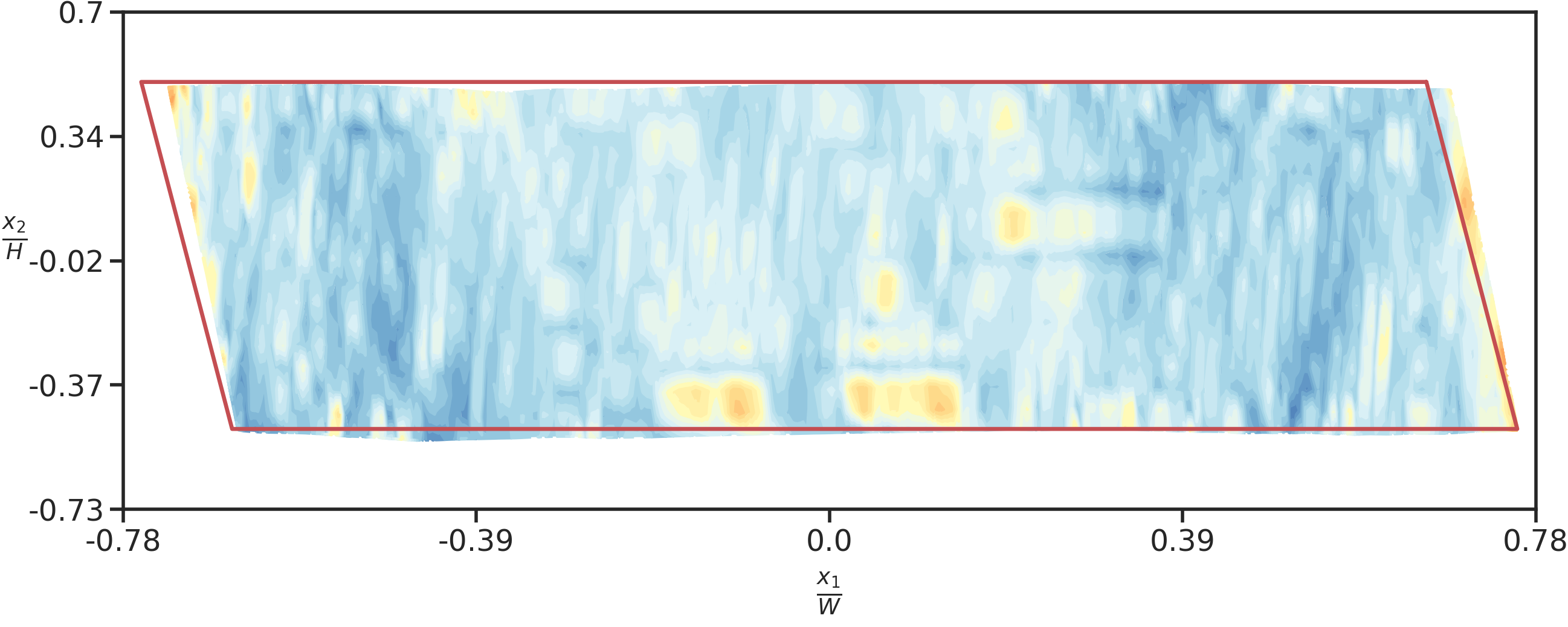}}
\subfloat[][$H = 0.5 \: \mu m$ (free-standing film)]{
\includegraphics[width=0.45\textwidth]{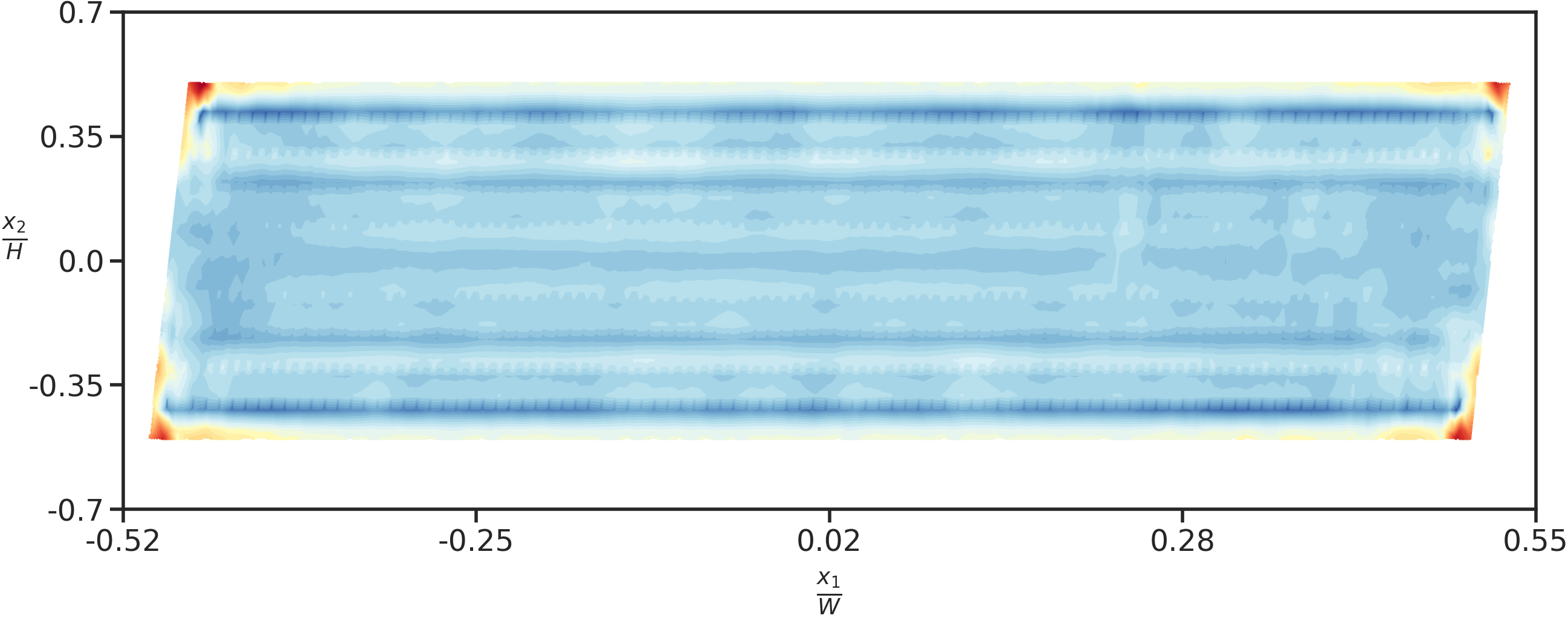}}
\quad
\subfloat{
\includegraphics[width = 0.45\textwidth]{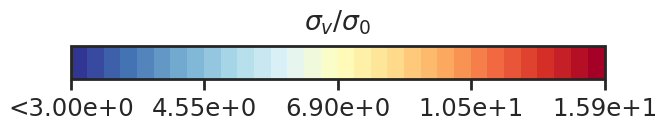}}
\caption{von Mises stress at $\Gamma_m = 30\%$ in the metal film for the micropillar sandwich with $45^{\circ}$ orientation and for the free-standing film. The corresponding $|e|$ for the micropillar is $1.15\%$, and the solid red line in (a) shows the boundary of the undeformed metal thin film.}
\label{fig:fig_J2_45}
\end{figure}

\begin{figure}[htbp]
\centering
\subfloat[][$H = 0.5 \: \mu m$ (micropillar)]{
\includegraphics[width=0.45\textwidth]{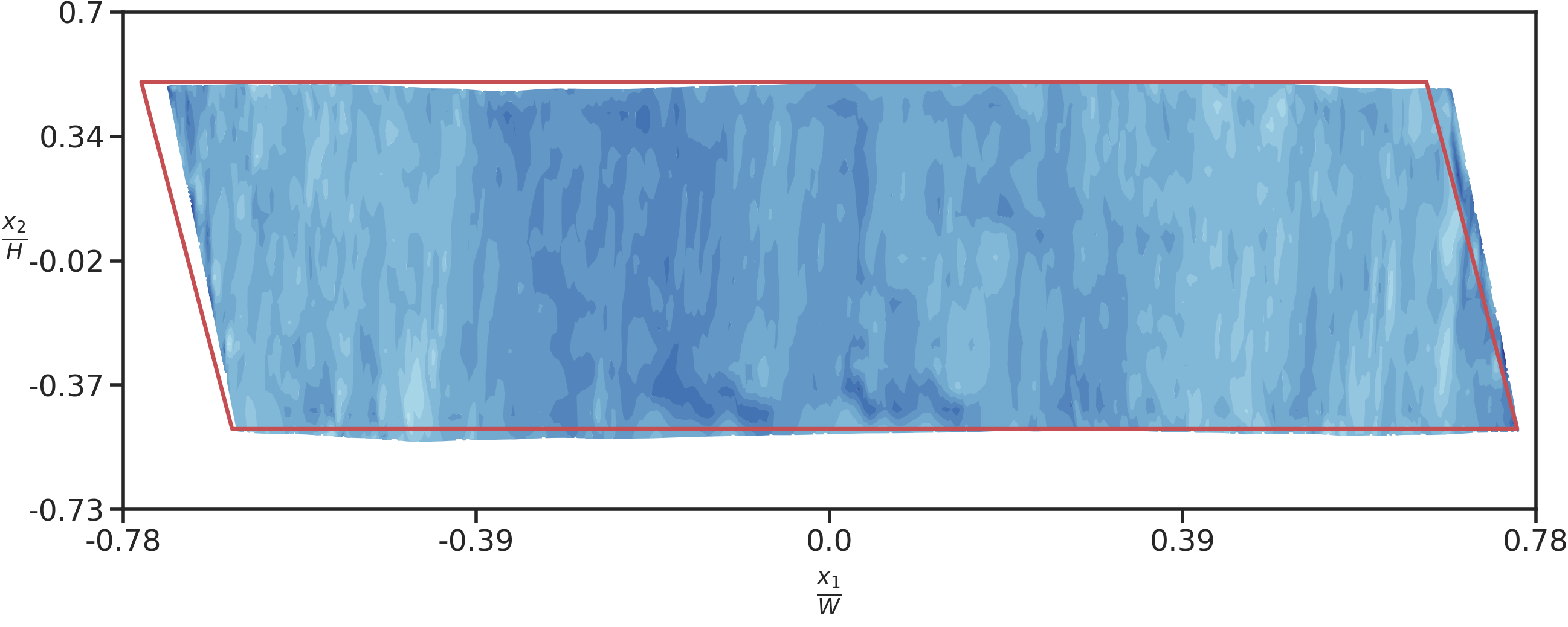}}
\subfloat[][$H = 0.5 \: \mu m$ (free-standing film)]{
\includegraphics[width=0.45\textwidth]{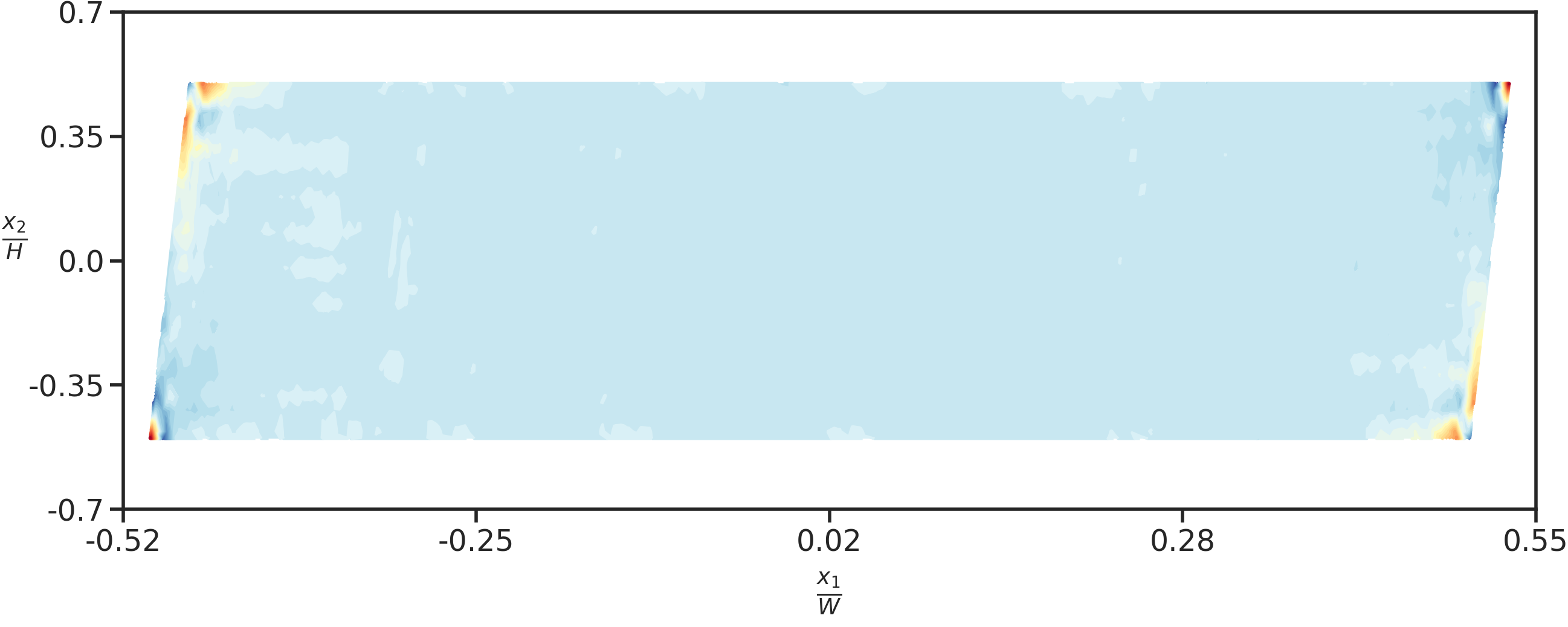}}
\quad
\subfloat{
\includegraphics[width = 0.45\textwidth]{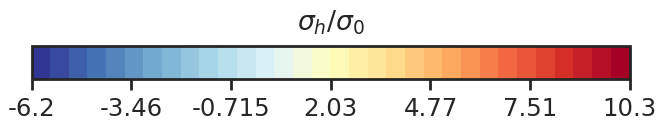}}
\caption{Hydrostatic stress at $\Gamma_m = 30\%$ in the metal film for the micropillar sandwich with $45^{\circ}$ orientation and for the free-standing film. The corresponding $|e|$ for the micropillar is $1.15\%$, and the solid red line in (a) shows the boundary of the undeformed metal thin film.}
\label{fig:fig_pressure_45}
\end{figure}

Fig.~\ref{fig:fig_rho_g_45_full} shows field plots of the norm of the Logarithmic strain tensor $\left( |ln (\bs{V})| \right)$ in the film within the micropillar, at $|e| = 3.5\%$ nominal compressive strain. As expected, for the smaller thickness, the magnitude of $|ln (\bs{V})|$ is higher as the nominal shear strain in the metal film is higher for smaller thickness (for equal lateral displacement).

Fig.~\ref{fig:fig_rho_g_45} shows the $\rho_g$ field plots at $\Gamma_m = 30\%$ for the free-standing film and the micropillar sandwich with $45^{\circ}$ orientation. As for the $90^{\circ}$ orientation case, there are no boundary layers in the film within the  micropillar here, while they are present for the free-standing film. 

Fig.~\ref{fig:fig_J2_45} and \ref{fig:fig_pressure_45} respectively show the von Mises stress $(\sigma_v)$ and the hydrostatic stress $(\sigma_h)$ field plots for the free-standing and micropillar sandwich configurations.


\subsection{Hydrostatic and normal stress in shear}
In small deformation elasticity or metal plasticity, shearing induced normal stresses are rarely observed. However, such couplings are observed in finite elastic \cite{poynting1909pressure} and plastic \cite{billington1977non} deformations. We explore such possibilities in our simulations of simple shearing of free-standing films and compression of micropillar sandwich with $45^{\circ}$ film orientation. To do so, we define the `driving' stress ($\sigma_2^{ap}$) in both configurations as the average reaction stress in the vertical direction on the top boundary of the domain. The averages of the various components of the Cauchy stress tensor over the metal film are also defined as ($\sigma_{1}^{avg}$, $\sigma_{2}^{avg}$, $\sigma_{3}^{avg}$, $\tau^{avg}$), denoting the averages of the ($T_{11}$, $T_{22}$, $T_{33}$, $T_{12}$) components, respectively. The averaged generated hydrostatic stress in the film is denoted by $\sigma_h^{avg}$.

The values of the driving stress and the averages of generated stress components for the micropillar sandwich and the free-standing film at $\Gamma_m = 30 \%$ are reported in Table \ref{tab:stresses}. The state of loading is multi-axial for both the micropillar sandwich and the free-standing film. However, the `average stress state' in the case of the micropillar sandwich is more hydrostatic (confined) in nature, while it is predominantly shear driven for the free-standing film, as also shown in Fig.~\ref{fig:fig_pressure_45}. The magnitude of hydrostatic stress in the free-standing film is much smaller than in the film in the micropillar, primarily due to the applied compressive loading on the pillar; nevertheless, the free-standing film generates compressive normal reactions at the boundary, for solely applied shear velocity boundary conditions (along with a constraint to motion of the boundary in the normal direction, of course).
\begin{table}[H]
    \centering
    \begin{tabular}{|c| c c c c c c| }\hline
        Specimen & $\sigma_{2}^{ap}/\sigma_0$  & $\sigma_{1}^{avg}/\sigma_0$
         & $\sigma_{2}^{avg}/\sigma_0$  & $\sigma_{3}^{avg}/\sigma_0$ & $\tau^{avg}/\sigma_0$ & $\sigma_h^{avg}/\sigma_0$ \\ \hline
        Micropillar sandwich & -6.086 & -0.122 & -6.241 & -2.283 & 0.109 & -2.882 \\ \hline
        Free-standing film & -0.048 & -0.01 & -0.035 & -0.026 & 2.912 & -0.024  \\ \hline
    \end{tabular}
    \caption{Driving stress and averaged stress components in the micropillar sandwich and the free-standing film at $\Gamma_m = 30 \%$.}
    \label{tab:stresses}
\end{table}
\noindent Despite this significant difference, the averaged hydrostatic stress normalized by the driving normal stress is of the same order of magnitude for both cases:
\begin{subequations}\label{eq:hydro_stress_data}
\begin{align}
\sigma_h^{avg}/\sigma_{2}^{ap} & = 0.474, \quad  \text{(Micropillar Sandwich)}\\
\sigma_h^{avg}/\sigma_{2}^{ap} & = 0.497 \quad \text{(Free-Standing Film)}
\end{align}
\end{subequations}
 (considering a normalization by applied stress component normal to the film in the micropillar increases the result by a factor of 2). This suggests a unifying `collapse' of data for understanding normal stress effects in nominal simple shearing, produced by significantly different applied loading conditions.

\begin{figure}[htbp]
    \centering
    \includegraphics[scale=0.35]{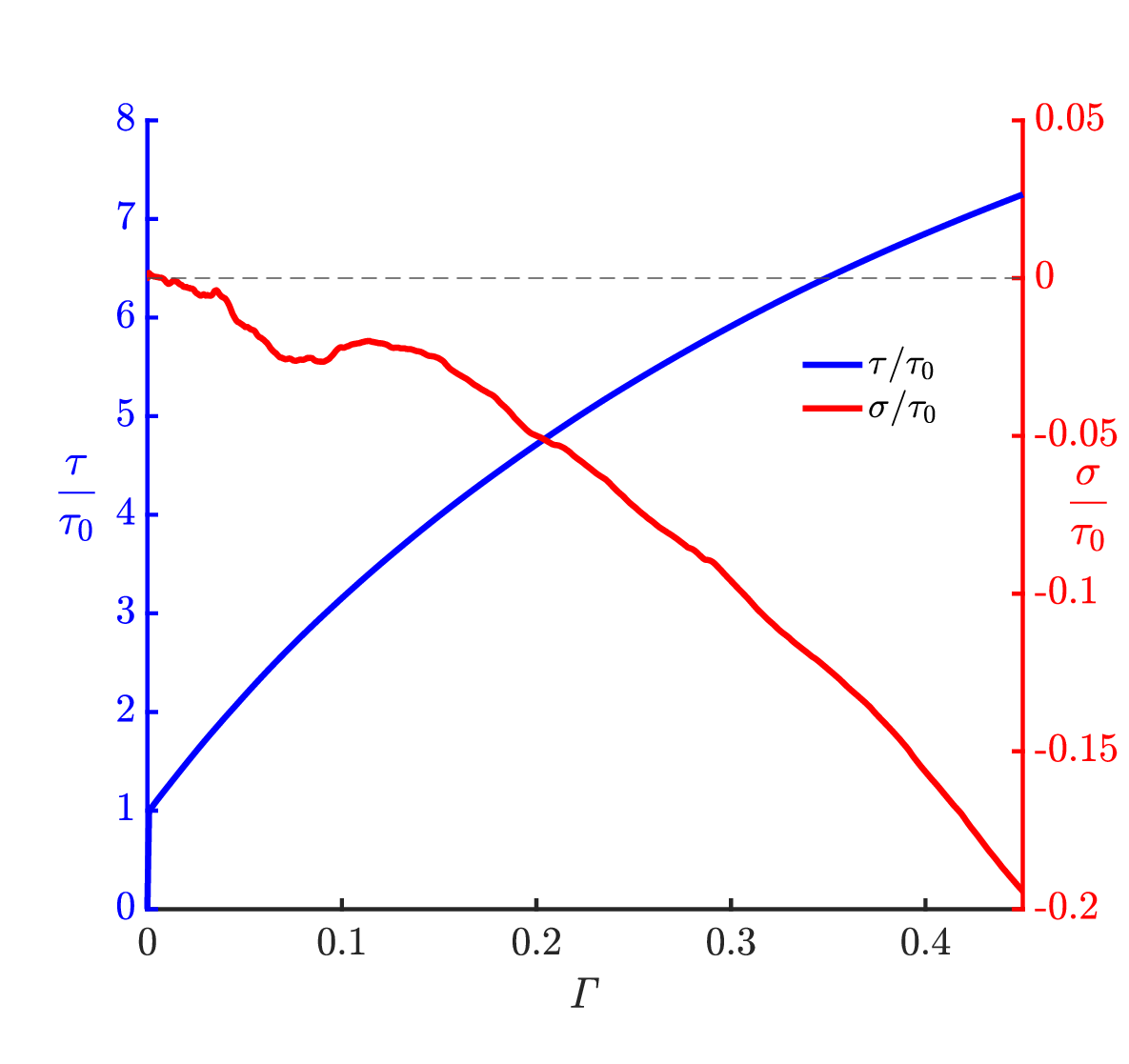}
    \caption{Normal and transverse reaction stress on the top boundary of the free-standing film subjected to simple shearing, for $H = 0.5 \: \mu m$.}
    \label{fig:normal_stress_effect}
\end{figure}


The normal stress observed here arises as a combination of the Swift effect \cite{billington1977non} and the Poynting effect \cite{poynting1909pressure}. One way to think about how these normal stresses are generated in the free-standing film is as follows: the $T_{22}$ component of the Cauchy stress tensor is non-zero, a nonlinear elastic Poynting effect to begin with. 
This further generates $T^{'}_{22}$ and $T^{'}_{11}$ components of the deviatoric stress tensor, and these components of stress generate plastic straining in the $L^p_{11}$ and $L^p_{22}$ components. However, due to the deformation constraints imposed by the simple shear boundary conditions in the $x_1$-direction on the lateral sides of the specimen, and the $x_2$-direction on the top and bottom of the domain, further $T_{11}$ and $T_{22}$ components of stress are generated. All of this combined generates a non-zero hydrostatic stress field and the idealized Swift effect shown in Fig.~\ref{fig:normal_stress_effect}, in simple shearing of the free-standing films. 

\subsection{Effect of b.c.s on plastic flow in the micropillar sandwich}

\begin{figure}[htbp]
    \centering
    \includegraphics[scale=0.35]{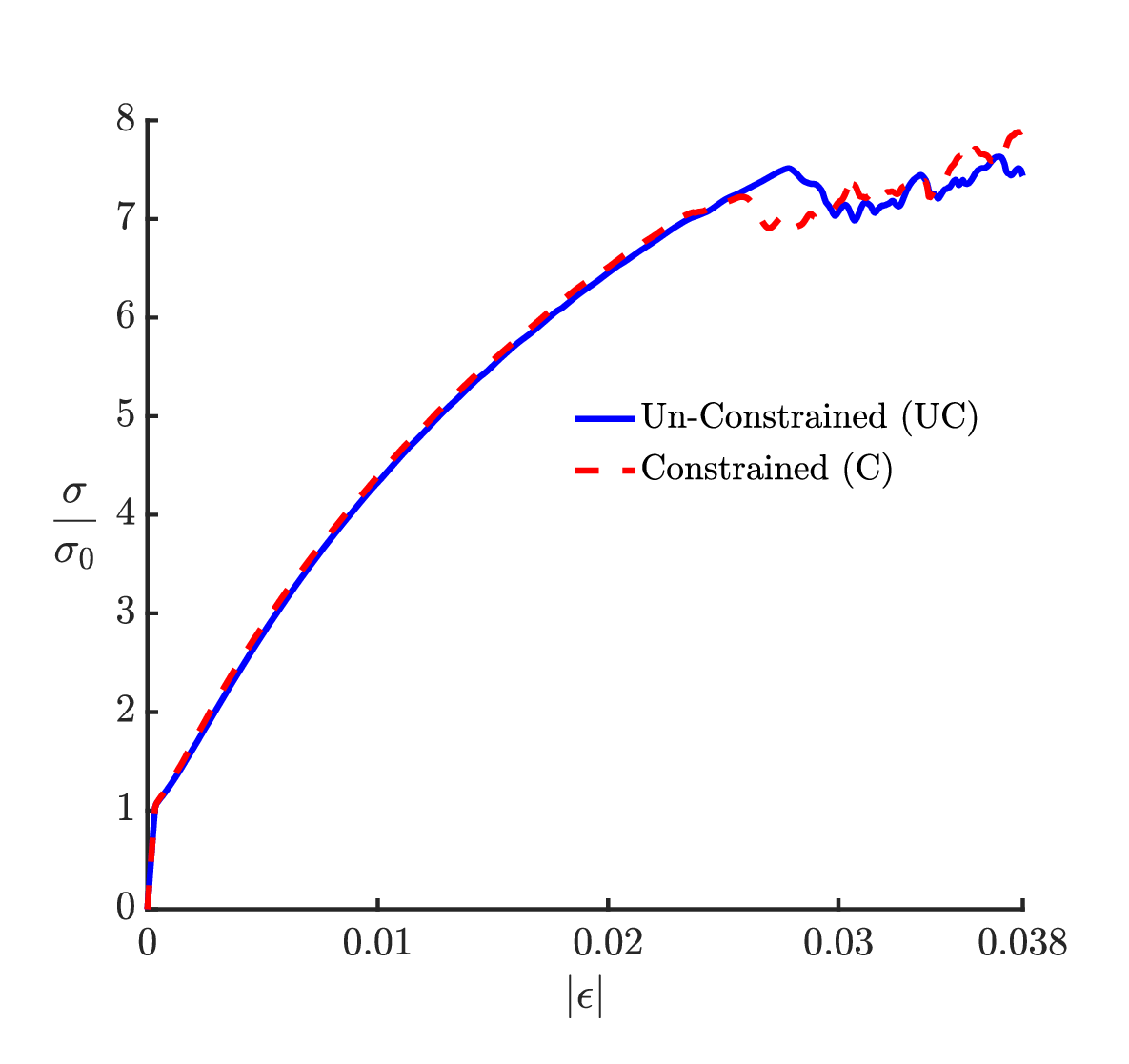}
    \caption{Stress-strain curve for the micropillar sandwich with $45^{\circ}$ orientation for $H = 0.8 \: \mu m$ case and with C and UC metal-ceramic interface.}
    \label{fig:fig_SS_45_0_8}
\end{figure}

\begin{figure}[H]
\centering
\subfloat[][UC]{
\includegraphics[width=0.45\textwidth]{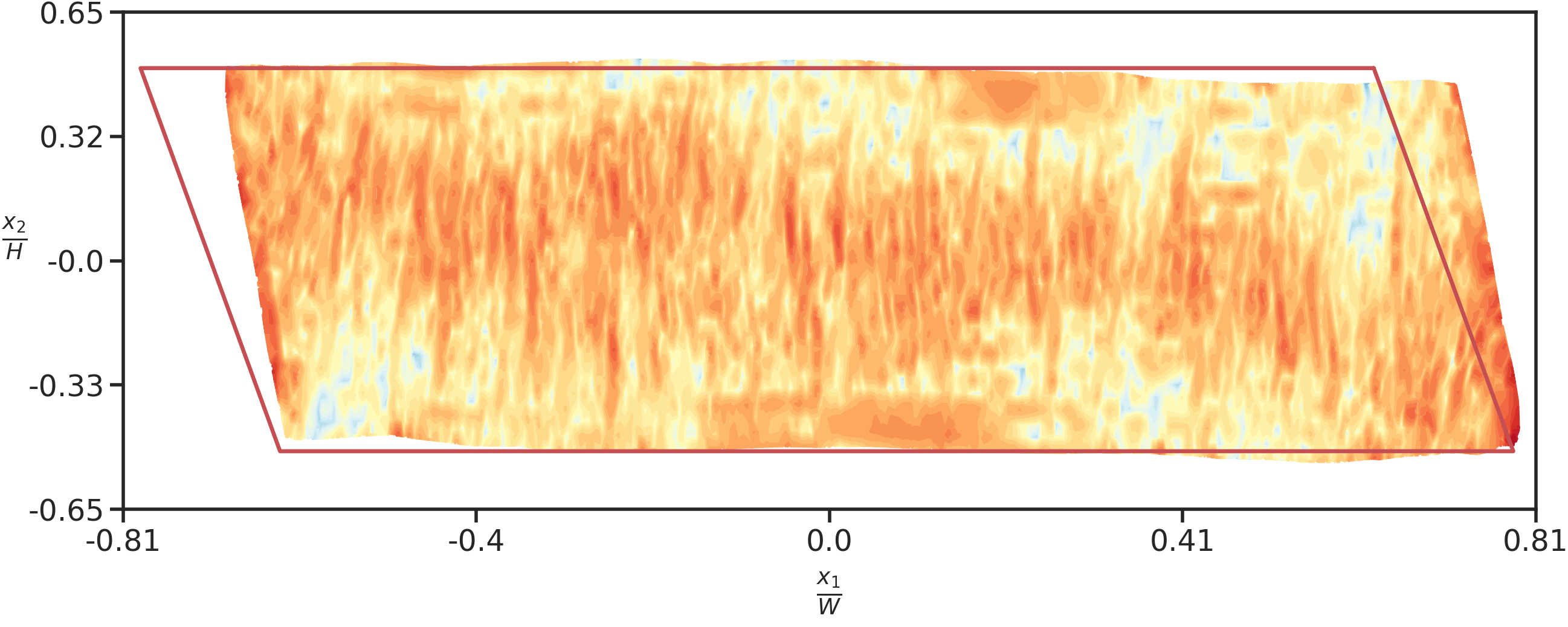}}
\subfloat[][C]{
\includegraphics[width=0.45\textwidth]{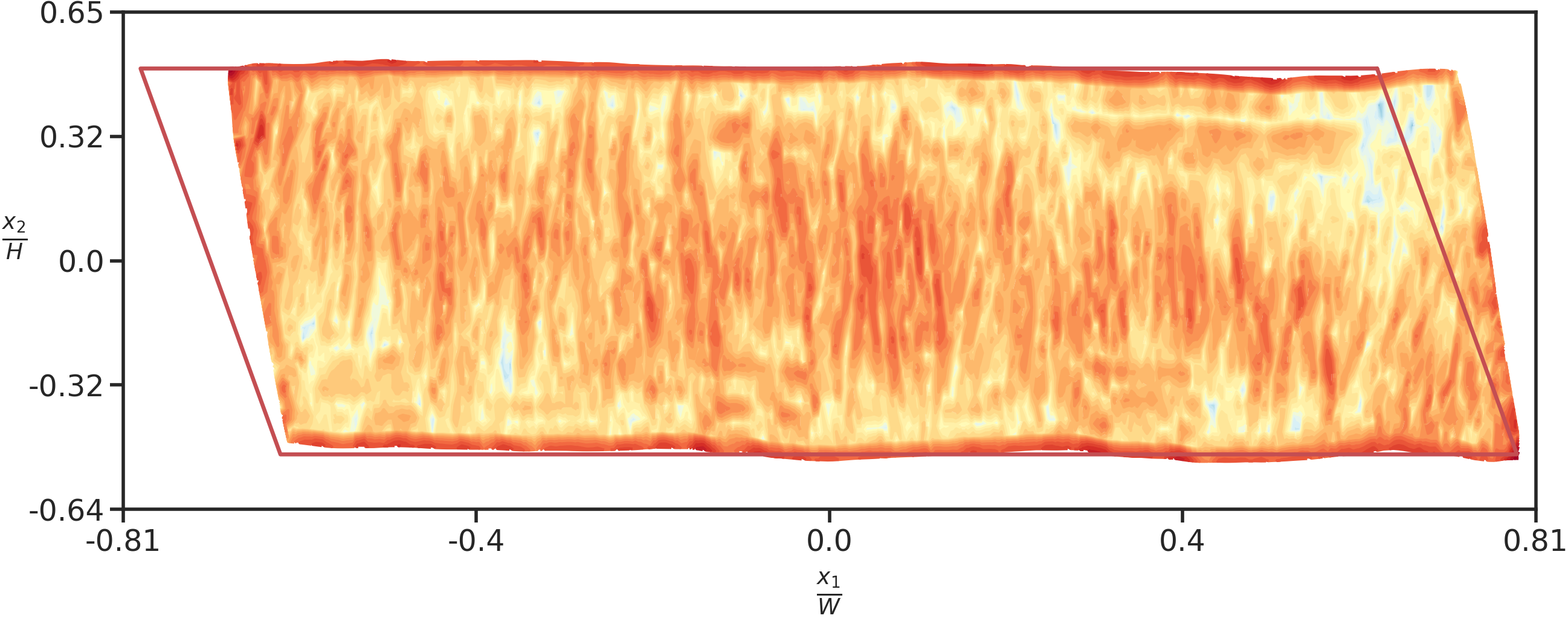}}
\quad
\subfloat{
\includegraphics[width = 0.45\textwidth]{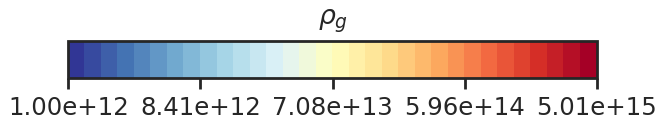}}
\caption{$\rho_g  = |\bs{\alpha}|/b \: (m^{-2})$ at $|e| = 3.5\%$ for $H = 0.8 \mu m$ with C and UC metal-ceramic interface. The solid red line shows the boundary of the undeformed metal thin film.}
\label{fig:fig_rho_g_45_0_8}
\end{figure}

Fig.~\ref{fig:fig_SS_45_0_8} shows the stress-strain curve and Fig.~\ref{fig:fig_rho_g_45_0_8} shows the $(\rho_g )$ field plots at $|e| = 3.5 \%$, for $H= 0.8 \: \mu m$, and for both the plastically constrained (C) and unconstrained (UC) metal-ceramic interfaces. The stress-strain response for the constrained case is marginally harder as compared to the unconstrained case. For the constrained case, there is a thin boundary layer in $\rho_g$ field at the metal-ceramic interface. The effect of the interface condition is not prominent here due to strong heterogeneity in plastic flow across the interface, and hence, having an elastic-plastic interface with plastic flow unconstrained has more or less the same effect as a plastically constrained interface. This is borne out also in the simple shear loading of free-standing films, where the effect of constrained/unconstrained boundary condition is much more prominent in stress-strain response, as shown by \cite{arora2020unification} (refer to Fig.\ 6(b) in their paper), due to the absence of constraining elastic material. 

Similar effects were observed in the study of mechanical response of multi-crystalline thin films in \cite{puri2011mechanical}. It was observed there that the effect of  constrained/unconstrained grain boundary conditions on the stress-strain response decreased on increasing the misorientation between adjacent grains, as the plastic flow through a grain boundary decreases with increase in the misorientation.



\section{Conclusion} \label{conclusion}

\noindent We have reported the first successful mechanistic understanding, through computational modeling, of experimentally observed effects of orientation in size effects of micropillar confined metal thin films undergoing large plastic deformations.  The results reported in the work of Mu et al.~\cite{mu2016dependence} form the motivation and experimental basis of our work.

Our contributions to the modeling and mechanistic understanding of such size effects in micropillar confined metal thin films are as follows:

\begin{itemize}

    \item The experimentally observed size effects obtained for the micropillar sandwich with thin films in two different orientations are dramatically different. Our simulations reproduce such size effects, and we provide a simple mechanistic explanation of why this must be the case. The size effect in the micropillar with a $90^{\circ}$ oriented film (compression) is stronger, as compared to the one with a $45^{\circ}$ oriented film (shearing), due to the stronger lateral constraint to material deformation imposed by the metal-ceramic interface for the compression case, which in turn causes more inhomogeneous deformation in the entire bulk of the film. This produces gradients in continued plastic straining which leads to more GND density and more hardening. In contrast, for the nominally simple-sheared film, whether in the micropillar configuration or free-standing, neither is a strong lateral constraint on material deformation available (just by geometrically intuitive reasons), nor is a constraint from the imposition of no plastic flow boundary conditions, as explained in Sec.~\ref{size_effects_section}. Hence, very modest size effects are observed in simple shear. Moreover, any theory that does not incorporate this geometric fact in the imposition of plastic flow b.c.s~ cannot differentiate between the differing constraints under direct compression and simple shearing, such plastic flow b.c.~constraints being one of the sources for the differing observed size effects in overall pillar compression with films in $45^\circ$ and $90^\circ$ orientations (to loading axis).
    
    The above lends valuable insight and understanding, of both scientific and technological value, into the mechanisms of size-dependent, large deformation plasticity at small scales and at engineering time-scales, not obtained by any other efforts known to us.
    
    In the current state-of-the-art of SGP theories, in one approach a threshold switch is introduced depending on the magnitude of plastic strain gradient at boundaries, based on which plastic straining is disallowed/allowed at the boundaries. Such a threshold is then fitted to experimental data for this set of experiments, but it is not clear what the microstructural justification of such a device might be, how to employ such a threshold under universal circumstances, and whether it breaks agreement of the theory with other experimental size-effect results in mesoscale plasticity. In another approach, a fractional SGP deformation theory of plasticity is proposed where the burden of prediction is left to the fitting of a new parameter of the theory whose microstructural origins is left unspecified.
    
     In contrast, no modification is made to the structure of the theory in our work, while being in good agreement with experiment and providing simple mechanistic understanding of a complex phenomena.
     
     \item We demonstrate the Swift and Poynting effects in our simulations and provide a mechanistic understanding of it.
    
    \item We make predictions of failure modes from our simulations for the micropillar sandwich in both orientations, and these are similar to those observed in the experimental works of \cite{mu2016dependence} and \cite{zhang2017mechanical}.
    
\end{itemize}

The results obtained in this paper further strengthens the case, beyond \cite{roy2006size, zhang2015single, puri2011mechanical, arora2020dislocation, arora2020unification, arora2020finite}, for MFDM as an appropriate model for dislocation mediated mesoscale plasticity. A shortcoming that needs to be addressed is to improve our model for plastic straining due to statistical dislocations, $\bfL^p$, so that its effect diminishes at smaller length scales.  

\section*{Acknowledgments}
This work was supported by the grant NSF OIA-DMR $\#2021019$.

\bibliographystyle{alpha}\bibliography{paper_template.bib}
\end{document}